\documentclass[aps,preprintnumbers,superscriptaddress,onecolumn,amsmath,amssymb,floatfix,pra]{revtex4-1}

\pdfoutput=1
\usepackage{graphicx}
\usepackage[usenames,dvipsnames]{xcolor}
\usepackage[colorlinks,bookmarks=false,citecolor=NavyBlue,linkcolor=Red,urlcolor=blue]{hyperref}
\usepackage[export]{adjustbox}

\usepackage{mathtools}

\usepackage{tikz}

\newcommand\be{\begin{equation}}
\newcommand\bea{\begin{eqnarray}}
\newcommand\bes{\begin{subequations}}
\newcommand\esu{\end{subequations}}
\newcommand\ee{\end{equation}}
\newcommand\eea{\end{eqnarray}}

\newcommand{\cmmnt}[1]{}

\newcommand\ba         {\begin{eqnarray} } 
\newcommand\ea         {\end{eqnarray} }

\def\doi{http://dx.doi.org/}

\usepackage{braket}

\newcommand{\dd}{{\rm d}}
\newcommand{\bv}{\textbf{v}}
\newcommand{\bu}{\textbf{u}}

\usepackage{dsfont}

\def\doi{http://dx.doi.org/}

\begin{document}

\title{Entanglement entropies of inhomogeneous Luttinger liquids}
\author{Alvise Bastianello}
\affiliation{Institute for Theoretical Physics, University of  Amsterdam, Science Park 904, 1098 XH Amsterdam, The  Netherlands}
\author{J\'er\^ome Dubail}
\affiliation{Laboratoire  de  Physique  et  Chimie  Th\'eoriques, CNRS, UMR 7019, Universit\'e  de  Lorraine, 54506 Vandoeuvre-les-Nancy, France}
\author{Jean-Marie St\'ephan}
\affiliation{Univ Lyon, CNRS, Universit\'e Claude Bernard Lyon 1, UMR5208, Institut Camille Jordan, F-69622 Villeurbanne, France}

\begin{abstract}
We develop a general framework to compute the scaling of entanglement entropy in
inhomogeneous one-dimensional quantum systems belonging to the Luttinger liquid universality class.
While much insight has been gained in homogeneous systems by making use of conformal field theory techniques, our focus is
on systems for which the Luttinger parameter $K$ depends on position, and conformal
invariance is broken. An important point of our analysis is that contributions stemming from the
UV cutoff have to be treated very carefully, since they now depend on position. We show that such
terms can be removed either by considering regularized entropies specifically designed to do so, or
by tabulating numerically the cutoff, and reconstructing its contribution to the entropy through the
local density approximation. We check our method numerically in the spin-1/2 XXZ spin chain in
a spatially varying magnetic field, and find excellent agreement.
\end{abstract}

\date{\today}


\maketitle

\section{Introduction}

Many quantum critical systems in one-dimension (1d), including various quantum gases or spin chains, belong to the universality class of Luttinger liquids \cite{haldane1981demonstration,haldane1981luttinger,haldane1981effective,cazalilla2004bosonizing,giamarchi2004quantum,tsvelik2007quantum,cazalilla2011one}. From an effective field theory perspective, Luttinger liquids are the one-dimensional systems whose low-energy description is provided by a massless free boson field theory (FT) in 1+1d, or in other words a U(1) conformal field theory~\cite{belavin1984infinite}. They are ubiquitous in the 1d quantum realm because ${\rm U}(1)$ symmetry itself is ubiquitous ---it is associated to particle number conservation, often naturally present in spin chains or quantum gases---, and because of the limited amount of relevant perturbations ---often forbidden by symmetry--- that could drive the system away from the renormalization group massless fixed point.

One celebrated result about Luttinger liquids is that the entanglement entropy of a subsystem $I = [x_1,x_2]$ in the ground state of an infinite homogeneous (i.e. translation invariant) system grows logarithmically with the size of the interval $I$ \cite{callan1994geometric,holzhey1994geometric,vidal2003entanglement,calabrese2004entanglement}. Specifically, for the entanglement entropy with R\'enyi index $\alpha$, the result reads \cite{calabrese2004entanglement}
\begin{equation}
	\label{eq:EE_CFT_renyi}
	S_\alpha (x_1,x_2) \, = \, \frac{1+1/\alpha}{6} \log \left( \frac{|x_1-x_2|}{a_\alpha}  \right)  ,
\end{equation}
up to corrections which go to zero as the length $|x_1-x_2|$ goes to infinity. Recall that the R\'enyi entanglement entropy of a pure state $\ket{\psi}$, which will always be the ground state in this paper, is defined by
\begin{equation}
 S_\alpha=\frac{1}{1-\alpha}\log \left(\textrm{Tr}\, \hat{\rho}_I^\alpha \right),
\end{equation}
where $\hat{\rho}_I=\textrm{Tr}_{\bar{I}} \ket{\psi}\bra{\psi}$ is the reduced density matrix for subsystem $I$, and $\bar{I}$ is its complement.
The constant $a_\alpha$ in (\ref{eq:EE_CFT_renyi}) is non universal. It cannot be fixed by global conformal invariance and its analytic determination can be extremely cumbersome even in free systems \cite{jin2004quantum,its2005entanglement,keating2004random,keating2005entanglement}. When comparing formula (\ref{eq:EE_CFT_renyi}) with numerical results in concrete microscopic systems, $a_\alpha$ is often treated as a fitting parameter.

Motivated by practical aspects of one-dimensional quantum gases in non-uniform trapping potentials \cite{gangardt2003stability,cazalilla2004bosonizing,petrov2004low,cazalilla2011one,wendenbaum2013hydrodynamic,dubail2017conformal,brun2017one,dubail2017emergence,brun2018inhomogeneous} or quantum spin chains in non-uniform magnetic fields \cite{balakrishnan1982inhomogeneous,eisler2009entanglement,lancaster2010quantum,landi2014flux,eisler2017front}, the purpose of this paper is to revisit the entanglement entropy of Luttinger liquids in the context of {\it inhomogeneous} 1d quantum systems, which are currently attracting a lot of attention, see e.g.  Refs.~\cite{Sinesquare,Katsura_2011,bertini2016transport,castro2016emergent,Sinesquare_ludwig,dubail2017conformal,eisler2017front,rodriguez2017more,tonni2018entanglement,murciano2018inhomogeneous,kulkarni2018quantum,dean2019nonequilibrium,stephan2019,ruggiero2019conformal,PhysRevLett.123.130602,QGHD} for recent works in that direction. 
We focus on {\it inhomogeneous Luttinger liquids}, namely on quantum systems whose effective field theory description corresponds to a free boson theory in a {\it curved metric} and with a {\it spatially varying} coupling strength or Luttinger parameter $K$ \cite{safi1995transport,maslov1995landauer,cazalilla2004bosonizing,dubail2017conformal,dubail2017emergence,eisler2017front,brun2018inhomogeneous}; the theory will be briefly reviewed in Sec. \ref{sec_luttinger}. 

In an inhomogeneous quantum system, we expect the entanglement entropy of an interval $I = [x_1,x_2]$ ---assumed to be larger than any microscopic length scale such as lattice spacing or interparticle distance--- to be of the form
\be
\label{eq_scal_ent}
S_\alpha(x_1,x_2)=S_\alpha^{{\rm FT}}(x_1,x_2)+c_\alpha(x_1)+c_\alpha(x_2), 
\ee
where $S^{{\rm FT}}(x,y)$ is the entanglement entropy calculated in the field theory with a certain UV cutoff ---that cutoff is discussed below---, and $c_\alpha(x), c_\alpha (y)$ are the microscopic contributions to the entanglement entropy around the two points $x$ and $y$. In homogeneous systems with entanglement entropy given by Eq. \eqref{eq:EE_CFT_renyi}, $c_\alpha(x)$ is independent of $x$. It can be thought of as $\frac{1+1/\alpha}{12} \log (a^{{\rm FT}}_\alpha / a_\alpha)$ where $a^{{\rm FT}}_\alpha / a_\alpha$ is the ratio of the UV cutoff in the field theory and a microscopic length scale in the quantum system.

In writing Eq. \eqref{eq_scal_ent}, two crucial features must be stressed. First, two microscopic models sharing the same low energy description have the same contribution $S_\alpha^{{\rm FT}}(x,y)$ to their entanglement entropies. 
This observation is at the root of the method developed in this paper: instead of considering a complicated interacting quantum model, it is possible to directly start from its effective field theory description as an inhomogeneous free boson theory. However, in replacing the true microscopic quantum model by its effective low energy description, one is left with the difficult task of finding the position-dependent microscopic contribution $c_\alpha(x)$.

On the other hand, the second feature apparent from Eq. \eqref{eq_scal_ent} is that the microscopic contribution $c_\alpha(x)$ is completely determined by the system in proximity of the points $x_1$ and $x_2$. This offers at least two possibilities, both of which we explore in great detail in the rest of this paper.

The first possibility is to consider a suitable combination of the entanglement entropy such that only universal features remain. For instance, one can define the $\epsilon$-regularized entanglement entropy as
\be\label{eq_reg_en}
S^{(\epsilon)}_\alpha (x_1,x_2)=\frac{1}{2}\left[S_\alpha (x_1,x_2)+S_\alpha(x_1-\epsilon,x_2+\epsilon)-S_\alpha(x_1-\epsilon,x_1)-S_\alpha(x_2,x_2+\epsilon)\right] 
\ee
(for similar definitions which appeared in the literature, see for instance Refs.~\cite{bianchi2014entanglement,bianchi2015entanglement}).
The effect of this regularization is to replace the cutoff-dependent contribution to the entanglement entropy by a function of the length $\epsilon$ which is determined only by the universal field theory part. Indeed, since the length $\epsilon$ can be fixed arbitrarily, one can chose $\epsilon$ to be much larger than the microscopic scale, yet much smaller than the length scale of the inhomogeneity. Then $S^{(\epsilon)}_\alpha(x,y)$ is fixed solely by $S^{\rm FT}_\alpha (x_1,x_2)$, as can be seen by plugging Eq.~\eqref{eq_scal_ent} into Eq.~(\ref{eq_reg_en}). 

The second possibility is to numerically tabulate the values of $c_\alpha$ by studying the homogeneous problem for different values of the external parameters. The inhomogeneous system can then be addressed by promoting $c_\alpha$ to a function of the position $x$ through the Local Density Approximation (LDA). In this approach, a numerical study of the homogeneous system together with the low energy effective theory gives access to the full entanglement entropy \eqref{eq_scal_ent}.

Below, we provide extensive checks of these two possibilities, thus demonstrating that the inhomogeneous Luttinger liquid theory can efficiently capture the scaling part of the entanglement entropies even in interacting inhomogeneous systems. The lack of conformal symmetry in Luttinger liquids with spatially varying coupling constant \cite{brun2018inhomogeneous} makes it impossible to apply standard methods of conformal invariance and to reach nice closed analytical results. Nevertheless the gaussianity of the theory still leads to expressions for entanglement entropies that can be efficiently evaluated numerically. The method we develop here is versatile as it applies to all quantum systems described by (inhomogeneous) Luttinger liquid theory, some of which ---including the Lieb-Liniger model--- can be very difficult to simulate by other means in general.

Finally, it is important to stress that we are dealing with a free compact boson, and that the compactification has non-trivial consequences in the computation of entanglement entropies. This is well understood in the homogeneous case with periodic boundary conditions \cite{furukawa2009mutual,calabrese2009entanglement,Calabrese_2011}, however for open boundary conditions the entanglement entropy was studied only recently by one of us \cite{bastianello2019renyi}. In this paper we focus on open boundary conditions and make extensive use of the results of Ref.~\cite{bastianello2019renyi}.

The paper is organized as follows. In Sec. \ref{sec_luttinger} we briefly review the theory of the inhomogeneous free boson in 1d and explain how to calculate entanglement entropies starting from the inhomogeneous Luttinger liquid Hamiltonian. We apply this formalism in Secs.~\ref{sec_ent_epsilon}-\ref{sec_ent} to study the entanglement entropy of an inhomogeneous XXZ chain. We perform DMRG checks in the XXZ chain to validate our approach. In Sec.~\ref{sec_ent_epsilon} we present our results for the $\epsilon$-regularized entanglement entropy with R\'enyi index $\alpha=2$. Sec.~\ref{sec_ent} is focused on the von Neumann entanglement entropy ($\alpha = 1$) and R\'enyi entropy with index $\alpha=2$ of a half-chain using numerical tabulation of the cutoff-dependent contribution $c_\alpha (x)$.  We conclude in Sec. \ref{sec_concl}. The more technical aspects of our results are presented in full detail in the Appendices.

\section{Entanglement entropy in the inhomogeneous free boson theory}
\label{sec_luttinger}

The field theory that underlies inhomogeneous Luttinger liquids (see e.g.~\cite{safi1995transport,maslov1995landauer,gangardt2003stability,gangardt2004universal,cazalilla2004bosonizing,dubail2017conformal,dubail2017emergence,eisler2017front,brun2018inhomogeneous}) is the free massless boson in a  metric $g$ with a spatially-dependent coupling constant ---which is called the Luttinger parameter $K$ in this paper---, see the action (\ref{eq:action}) below. In this section we start by briefly reviewing this theory in Hamiltonian formalism (better suited than lagrangian formalism for the purposes of this paper). Then we explain how, discretizing the theory on a finite grid, one can obtain formulas that allow to calculate the entanglement entropy in terms of the Green's function of that free theory.

\subsection{Definition and Hamiltonian}

The Hamiltonian of a {\it homogeneous} Luttinger liquid defined on the interval $[0,L]$ is (we follow the notational conventions of the textbook of Giamarchi \cite{giamarchi2004quantum}):
\be\label{H_hom}
\hat{H}^{\text{LL}}=\frac{1}{2\pi}\int_0^L \dd x \, v\,\left[  K  ( \partial_x \hat{\theta})^2+\frac{1}{K} (\partial_x\hat{\phi} )^2\right] ,
\ee
where $v$ is the sound velocity in the fluid, $K$ is the Luttinger parameter which encodes the renormalized interaction strength between the constituents of the microscopic model, and $\hat{\phi}$ and $\pi\hat{\Pi}=  \partial_x \hat{\theta}$ are canonically conjugated bosonic fields,
\be
[\hat{\phi}(x),\hat{\Pi}(y)]=i\delta(x-y).
\ee
Throughout this paper, we focus on Dirichlet boundary conditions for the field $\hat{\phi}$ (as in Refs.~\cite{cazalilla2004bosonizing,brun2017one,brun2018inhomogeneous}),
\be\label{eq_OPC}
\hat{\phi}(x=0)=\hat{\phi}(x=L)=0\,.
\ee

Importantly, the fields $\hat{\phi}$, $\hat{\Pi}$ are compactified: $\hat{\theta}$ is identified with $\hat{\theta} + 2\pi$, while $\hat{\phi}$ is identified with $\hat{\phi} + \pi$. One way of understanding this is to look at the identification of operators in the microscopic theory with local operators in the field theory. For instance, in a microscopic model of bosons or fermions, the particle density operator has an expansion of the form $\hat{n}(x) \, = \, n_0 - \frac{1}{\pi} \partial_x \hat{\phi} (x) + \dots$,  where $n_0$ is the mean value of the density. Thus, adding or removing one particle corresponds to a $\pm \pi$-jump of the field $\hat{\phi}$. Such $\pm \pi$ jumps are most conveniently interpreted as non-zero windings when $\hat{\phi}$ takes values in the circle $\mathbb{R} / (\pi \mathbb{Z})$. Similarly, the field $\hat{\theta}$ is usually interpreted as a phase, and therefore takes values in the circle $\mathbb{R} / (2\pi \mathbb{Z})$. We will see in Sec.~\ref{sec:compactification} that this compactification induces non-trivial technicalities in the computation of entanglement entropies.

Finally, we stress that for microscopic models that can be mapped to free fermions, the Luttinger parameter $K$ is always equal to one (see for instance the discussions of that point in Refs.~\cite{dubail2017emergence,granet2018inhomogeneous}), but in the presence of interactions it can be any positive real number.

\vspace{0.4cm}

Now let us turn to the {\it inhomogeneous} case. In the limit of very slow variations of the mean density $n_0(x)$, it is still possible to describe the low-energy fluctuations of the system by the two canonically conjugated fields $\hat{\phi}(x)$ and $\hat{\Pi}(x)$, if one can identify the correct inhomogeneous generalization of the Hamiltonian \eqref{H_hom}. 
Of course, the inhomogeneous generalization must be such that, locally, it reproduces the results of the homogeneous model. The simplest way of ensuring this is to promote the parameters in the Hamiltonian \eqref{H_hom} to functions of the position $v\to v(x),$ $K\to K(x)$, the local values of which are fixed by the underlying microscopic theory through LDA \cite{cazalilla2004bosonizing,dubail2017conformal,dubail2017emergence,eisler2017front,brun2018inhomogeneous},
\be\label{H_inh}
\hat{H}^{\text{LL}}=\frac{1}{2\pi}\int_0^L \dd x \, v(x)\,\left[ K(x) ( \pi \hat{\Pi})^2+\frac{1}{K(x)} (\partial_x\hat{\phi})^2\right] .
\ee
Notice however that, as discussed in detail in Refs. \cite{dubail2017emergence,brun2018inhomogeneous,granet2018inhomogeneous}, promoting the model to inhomogeneous $v(x)$ and $K(x)$ leads to possible ambiguities. Different choices are in principle allowed: for example, substituting $1/K (\partial_x \hat{\phi})^2$ in the homogeneous Hamiltonian (\ref{H_hom}) by $[\partial_x (\sqrt{1/K(x)}\hat{\phi})]^2$ instead of $1/K(x)(\partial_x\hat{\phi})^2$ would lead to another inhomogeneous Hamiltonian which would look equally valid. In Refs.~\cite{brun2018inhomogeneous,granet2018inhomogeneous} it is properly argued that, in lagrangian formalism, the correct action reads (the field $h(x,\tau)$ in these references is $2\times \phi(x,\tau)$ here)
\begin{equation}
	\label{eq:action}
	\mathcal{S} \, = \, \frac{1}{2\pi} \int \frac{\sqrt{g} \dd^2 {\rm x} }{K({\rm x})} g^{a b} (\partial_a \phi) (\partial_b \phi) ,
\end{equation}
where ${\rm x} = ({\rm x}^1, {\rm x}^2) =(x, \tau)$, $\tau$ is the imaginary time coordinate, and the two-dimensional metric $g_{ab}$ has euclidean signature and reads in our case
\begin{equation}
	\dd s ^2 \, = \, g_{ab} \dd{\rm x}^a \dd {\rm x}^b \, = \, \dd x^2 + v(x)^2 \dd \tau^2 .
\end{equation}
In Appendix~\ref{app:S_H} we show that the Hamiltonian (\ref{H_inh}) is indeed the one that is obtained from the action (\ref{eq:action}). The covariant form of the action suggests the change of variables 
\be\label{eq_iso}
x \rightarrow \tilde{x} = \left(\int^L_0 \frac{\dd y}{v(y)}\right)^{-1}\int^x_0 \frac{\dd y}{v(y)},
\ee
which puts the metric in the form $\dd s^2  \, \propto \, \dd \tilde{x}^2 + \dd \tau^2$ \cite{dubail2017conformal} and $\tilde{x}\in[0,1]$. Hence the position-dependent sound velocity drops from most computations, and we find that it also facilitates numerical calculations (see Appendix~\ref{app_num_discretelutt}). The position-dependent Luttinger parameter $K(x)$, however, cannot be absorbed in a simple change of variables, and the ground state correlations can no longer be calculated analytically, contrary to the case when $K$ is a constant. This is an important complication inherent to interacting inhomogeneous Luttinger liquids. The ground state correlations can all be expressed in terms of the Green's function of a generalized laplacian in 2d, $\nabla_{\rm x} \frac{1}{K({\rm x})} \nabla_{\rm x}$, and the problem is formally analogous to calculating the Coulomb energy of distributions of electric and magnetic charges in a 2d electrostatic problem with a {\it spatially varying} dielectric constant $1/K({\rm x})$. We refer to Ref. \cite{brun2018inhomogeneous} for a more detailed discussion of ground state correlations in the inhomogeneous Luttinger liquid.

In this paper, we rather rely on a numerical evaluation of the ground state correlations $\left< \hat{\phi}(x) \hat{\phi} (x') \right>$, $\left< \hat{\phi}(x) \hat{\Pi} (x') \right>$, $\left< \hat{\Pi}(x) \hat{\Pi} (x') \right>$ which is based on a discretization of the system on a finite grid. We refer to Appendix \ref{app_num_discretelutt} for further details on our numerical procedure.

\subsection{Entanglement entropy of an interval at the edge}
\label{sub_ee_edge}

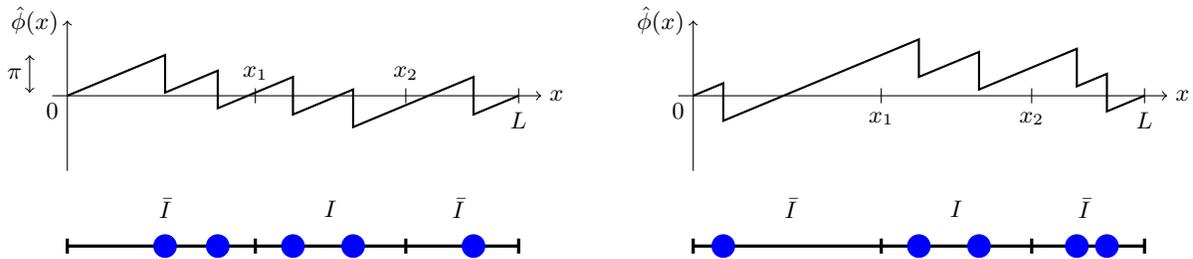
\begin{figure}[t]
	\begin{tikzpicture}
	\begin{scope}
		\draw[very thick] (0,0) -- (6,0);
		\draw[very thick] (0,-0.1) -- ++(0,0.2);
		\draw[very thick] (2.5,-0.1) -- ++(0,0.2);
		\draw[very thick] (4.5,-0.1) -- ++(0,0.2);
		\draw[very thick] (6,-0.1) -- ++(0,0.2);
		\filldraw[blue] (1.3,0) circle (1.5mm); 
		\filldraw[blue] (2,0) circle (1.5mm); 
		\filldraw[blue] (3,0) circle (1.5mm); 
		\filldraw[blue] (3.8,0) circle (1.5mm); 
		\filldraw[blue] (5.4,0) circle (1.5mm);
		\draw (1.3,0.5) node{$\bar{I}$};
		\draw (3.5,0.5) node{$I$};
		\draw (5.2,0.5) node{$\bar{I}$};
	\end{scope}
	\begin{scope}[yshift=1cm]
		\draw (2.5,1-0.1) -- ++(0,0.2) node[above]{$x_1$};
		\draw (4.5,1-0.1) -- ++(0,0.2) node[above]{$x_2$};
		\draw (6,1-0.1) node[below]{$L$} -- ++(0,0.2);
		\draw[->] (-0.2,1) -- (6.3,1) node[right]{$x$};
		\draw[->] (0,0) -- (0,2) node[left]{$\hat{\phi}(x)$};
		\draw[thick] (0,1) -- (1.3,1+0.5416) -- (1.3,0.5+0.5416) -- (2,0.5+0.8333) -- (2,0+0.8333) -- (3,0+1.25) -- (3,-0.5+1.25) -- (3.8,-0.5+1.583) -- (3.8,-1+1.583) -- (5.4,-1+2.25) -- (5.4,-1.5+2.25) -- (6,1);
		\draw[<->] (-0.5,1+0.54) -- (-0.5,1.04);
		\draw (-0.7,1.3) node{$\pi$};
		\draw (-0.2,0.8) node{$0$};
	\end{scope}
	\end{tikzpicture}
	\qquad
	\begin{tikzpicture}
	\begin{scope}
		\draw[very thick] (0,0) -- (6,0);
		\draw[very thick] (0,-0.1) -- ++(0,0.2);
		\draw[very thick] (2.5,-0.1) -- ++(0,0.2);
		\draw[very thick] (4.5,-0.1) -- ++(0,0.2);
		\draw[very thick] (6,-0.1) -- ++(0,0.2);
		\filldraw[blue] (0.4,0) circle (1.5mm); 
		\filldraw[blue] (3,0) circle (1.5mm); 
		\filldraw[blue] (3.8,0) circle (1.5mm); 
		\filldraw[blue] (5.1,0) circle (1.5mm);
		\filldraw[blue] (5.5,0) circle (1.5mm); 
		\draw (1.3,0.5) node{$\bar{I}$};
		\draw (3.5,0.5) node{$I$};
		\draw (5.2,0.5) node{$\bar{I}$};
	\end{scope}
	\begin{scope}[yshift=1cm]
		\draw (2.5,1-0.1) node[below]{$x_1$} -- ++(0,0.2);
		\draw (4.5,1-0.1) node[below]{$x_2$} -- ++(0,0.2);
		\draw (6,1-0.1) node[below]{$L$} -- ++(0,0.2);
		\draw[->] (-0.2,1) -- (6.3,1) node[right]{$x$};
		\draw[->] (0,0) -- (0,2) node[left]{$\hat{\phi}(x)$};
		\draw[thick] (0,1) -- (0.4,1+0.1666) -- (0.4,0.5+0.1666) -- (3,0.5+1.25) -- (3,0+1.25) -- (3.8,0+1.583) -- (3.8,-0.5+1.583) -- (5.1,-0.5+2.125) -- (5.1,-1+2.125) -- (5.5,-1+2.2916) -- (5.5,-1.5+2.2916) -- (6,1);
		\draw (-0.2,0.8) node{$0$};
	\end{scope}
	\end{tikzpicture}
	\caption{\emph{Cartoon of two configurations of particles in the microscopic model, and the corresponding expectation value of the field $\hat{\phi}(x)$, which jumps by $-\pi$ at the position of each particle according to $\hat{n}(x) = n_0(x)- \frac{1}{\pi} \partial_x \hat{\phi}$. Notice that the two configurations of particles are globally different, however they are {\it identical inside the subsystem $I$}. In terms of the field $\hat{\phi}(x)$, the two configurations inside the interval $I$ are viewed as different unless one identifies $\hat{\phi}(x)$ with $\hat{\phi}(x) + \pi$. In other words, if one simply treats the field $\hat{\phi}(x)$ as real-valued, then one is overcounting the configurations inside the interval $I$; if instead the field $\hat{\phi}(x)$ is viewed as taking values in $\mathbb{R}/(\pi \mathbb{Z})$ then it correctly encodes particle configurations. It is extremely important to properly take into account the compactification in the calculation of the entanglement entropy of an interval $I$ in the bulk. Notice that, because of the Dirichlet boundary condition satisfied by $\hat{\phi}(x)$, there is no problem of overcounting when the interval $I$ touches the boundary.}}
	\label{fig:twoconf}
\end{figure}

Let us consider a bipartition $I \cup \bar{I}$ of the system, with $I = [0, x_1]$ and $\bar{I} = [x_1,L]$. 
For this choice of bipartition, one does not need to be particularly careful about the compactification of the field $\hat{\phi}(x)$ (see Fig.~ \ref{fig:twoconf}). The calculation can be done straightforwardly, following for instance the review~\cite{peschel2009reduced}. Here we write the calculation for continuous fields, without paying attention to problems of UV regularization. In practice, the formulas we present here will be used in Sections \ref{sec_ent} and \ref{sec_ent_epsilon} by discretizing the system on a finite grid, thus naturally providing a UV regularization (see Appendix \ref{app_num_discretelutt} for details).

We define the bosonic annihilation/creation operators
\be
\hat{\varphi}(x)=\frac{n_0^{\frac{1}{2}}(x) \hat{\phi}(x)-i n_0^{-\frac{1}{2}}(x) \hat{\Pi}(x)}{\sqrt{2}}\, \hspace{2pc}\hat{\varphi}^\dagger(x)=\frac{n_0^{\frac{1}{2}}(x) \hat{\phi}(x)+i n_0^{-\frac{1}{2}}(x) \hat{\Pi}(x)}{\sqrt{2}}, 
\ee
such that $[\hat{\varphi}(x),\hat{\varphi}^\dagger(y)]=\delta(x-y)$. The mean density $n_0(x)$ appears because of dimensionality. We also define the correlation operator $C(x,y)$ in  block form
\be
C(x,y)=\begin{pmatrix}\langle\hat{\varphi}(x)\hat{\varphi}^\dagger(y)\rangle && \langle\hat{\varphi}^\dagger(x)\hat{\varphi}^\dagger(y)\rangle \\ \langle\hat{\varphi}(x)\hat{\varphi}(y)\rangle &&\langle \hat{\varphi}^\dagger(x)\hat{\varphi}(y)\rangle\end{pmatrix} ,
\ee
with the following notational convention for product of operators: $(A B)(x,y) = \int\dd z\, A(x,z)B(z,y)$.

Let us call $C_I$ the correlator restricted to the subsystem $I = [0,x_1]$. Since the correlation functions of the whole system satisfy Wick's theorem, correlation functions restricted to the subspace $I$ automatically satisfy Wick's theorem as well. This implies that the reduced density matrix $\hat{\rho}_I$ is gaussian in terms of the bosons and unambiguously fixed by the correlator $C_I$. In particular, there exists a canonical transformation to new bosonic fields $\hat{\gamma}(x)$, $\hat{\gamma}^\dagger(x)$
\be
\label{eq:Mdiag}
\begin{pmatrix} \hat{\varphi}(x) \\ \hat{\varphi}^\dagger(x)\end{pmatrix}=\int \dd k \, M(x,k)\begin{pmatrix} \hat{\gamma}(k) \\ \hat{\gamma}^\dagger(k)\end{pmatrix} ,
\ee
which diagonalizes the correlation operator:
\be
\label{eq:canonicalC}
(M C_I M^\dagger)(k,q)= \begin{pmatrix}\langle\hat{\gamma}(k)\hat{\gamma}^\dagger(q)\rangle && \langle\hat{\gamma}^\dagger(k)\hat{\gamma}^\dagger(q)\rangle \\ \langle\hat{\gamma}(k)\hat{\gamma}(q)\rangle &&\langle \hat{\gamma}^\dagger(k)\hat{\gamma}(q)\rangle\end{pmatrix} = \delta(k-q)\begin{pmatrix}1+\frac{1}{e^{\lambda(k)}-1} && 0\\ 0 && \frac{1}{e^{\lambda(k)}-1} \end{pmatrix}\, .
\ee
This transformation also diagonalizes the reduced density matrix,
\be
\label{eq:canonical}
\hat{\rho}_I \, = \, \frac{ e^{-\int \dd k \, \lambda(k)\hat{\gamma}^\dagger(k)\hat{\gamma}(k)}}{{\rm Tr}\, e^{-\int \dd k \, \lambda(k)\hat{\gamma}^\dagger(k)\hat{\gamma}(k)}}\, .
\ee
The entanglement entropy is then related to the eigenvalues $\lambda (k)$:
\be
\label{eq_freeb_renyi}
 S_\alpha = \frac{1}{1-\alpha} \log [ {\rm Tr} \, \hat{\rho}_I^\alpha ] = \frac{1}{1-\alpha}\int \dd k\, \log\left[\frac{(1-e^{-\lambda(k)})^\alpha}{1-e^{-\alpha \lambda(k)}}\right]  =\int \dd k\, f_\alpha (\lambda(k)) ,
\ee
where $f_\alpha (\lambda) =  \frac{1}{1-\alpha} \log \left[\frac{(1-e^{-\lambda})^\alpha}{1-e^{-\alpha \lambda}} \right]$ and, in the case $\alpha =1$, $f_1 (\lambda) = \lim_{\alpha \rightarrow 1} f_\alpha (\lambda) =  \frac{\lambda}{e^\lambda-1}-\log\left(1-e^{-\lambda}\right)$.

We see that the non-trivial part of the calculation is the diagonalization of the correlation operator $C_I$, Eqs. (\ref{eq:Mdiag})-(\ref{eq:canonicalC}). Let us elaborate on that point. In order to preserve the canonical commutation rules, the change of basis (\ref{eq:canonical}) must be symplectic,
\be
M\sigma M^\dagger=\sigma \qquad \quad {\rm with} \qquad \quad \sigma(k,q) \, = \, \delta(k-q)\begin{pmatrix}1 && 0\\ 0 && -1 \end{pmatrix} .
\ee
It is convenient to take advantage of the symplectic transformation to observe that $M^\dagger = \sigma M^{-1} \sigma$, so
\be
(MC_I \sigma M^{-1})(k,q)= (MC_I M^\dagger \sigma)(k,q) =  \delta(k-q)\begin{pmatrix}1+\frac{1}{e^{\lambda(k)}-1} && 0\\ 0 && -\frac{1}{e^{\lambda(k)}-1} \end{pmatrix}\, .
\ee
Thus, the spectrum of $C_I \sigma$ gives access to $\lambda(k)$, which then gives the entanglement entropy through Eq. \eqref{eq_freeb_renyi}. Again, we emphasize that, in practice, we discretize the inhomogeneous Luttinger liquid on a finite grid (Appendix \ref{app_num_discretelutt}), so the correlation matrix $C_I$ is a finite-size square matrix and the spectrum of $C_I \sigma$ is obtained numerically.

\subsection{Entanglement entropy of an interval in the bulk: the role of compactification}
\label{sec:compactification}

Next, we turn to the case of a bipartition $I \cup \bar{I}$ with $I = [x_1,x_2]$ and $\bar{I} = [0, x_1] \cup [x_2,L]$, with $0<x_1<x_2<L$. This situation is more complicated than the one in the previous section because of the compactification of the field $\hat{\phi}$ which now plays a role. To see this, consider the two configurations of particles in the microscopic model drawn in Fig.~\ref{fig:twoconf}. Although they are identical inside the interval $I$, the field $\hat{\phi}(x)$ takes different real values inside $I$. One sees that the configurations are identical only if one identifies $\hat{\phi}(x)$ with $\hat{\phi}(x) + \pi$. Not taking this compactification into account would lead to wrong results for the entanglement entropy.

It is well known that dealing with compactification in the context of calculating entanglement entropies is not an easy task. Indeed, the same problem is faced when calculating the entanglement entropy of a multi-interval in an infinitely long system \cite{furukawa2009mutual,calabrese2009entanglement,Calabrese_2011,dupic2018entanglement}, and there only the R\'enyi entropies with integer $\alpha \geq 2$ have been accessed through a direct analytical calculation.  No analytical continuation to continuous R\'enyi parameter $\alpha$ has been found so far even in that well-studied case, albeit the analytical continuation can sometimes be performed numerically \cite{PhysRevD.89.025018,De_Nobili_2015}. In Ref. \cite{Ruggiero_2018,Rajabpour_2012}, by means of conformal blocks techniques, the R\'enyi entropies for disjoint intervals were analyzed in the PBC case, resulting in an expansion in the small distance among the two intervals: such an expression can be analytically continued term by term, giving access to the same expansion for the Von Neumann entropy. However, connecting this method with the exact results of Ref. \cite{furukawa2009mutual,calabrese2009entanglement,Calabrese_2011,dupic2018entanglement}
is still an open challenge.

Therefore, in this section we also restrict ourselves to the case of R\'enyi index $\alpha$ integer and $\geq 2$. The calculation leading to this result was done very recently by one of us in Ref.~\cite{bastianello2019renyi}, and we refer to that reference for details of the derivation. The result of Ref.~\cite{bastianello2019renyi} reads
\be\label{eq_reny_int}
S_\alpha (x_1,x_2)=\frac{1}{1-\alpha}\log\left[\prod_{a=1}^{\alpha-1}\sqrt{\frac{1}{\pi^3\mathcal{I}_{a/\alpha}}\frac{1}{\det\left(\frac{1+\Phi^{-1}\Phi_{a/\alpha}}{2}\right)}}\sum_{\{m_j\}_{j=1}^{\alpha-1}}  \exp\Bigg\{-4\sum_{a,b=1}^{\alpha-1}\mathcal{M}_{ab}\, m_a m_b\Bigg\}\, \right]+ C_\alpha\, ,
\ee
where the constant offset does not depend on $x_1$, $x_2$, but may depend on $\alpha$. The summation is over $m_1, m_2, \dots, m_{\alpha-1} \in \mathbb{Z}$. The matrix $\mathcal{M}$ is a square matrix of size $(\alpha-1) \times (\alpha-1)$, while $\Phi$ and $\Phi_\omega$ are operators: their determinants are computed by means of a suitable discretization of the continuum model on a finite grid, before taking the limit of zero lattice spacing. In doing so, the entanglement entropy is well defined except for a constant offset, which diverges logarithmically in the lattice size \cite{bastianello2019renyi}, in agreement with the homogeneous result \eqref{eq:EE_CFT_renyi}.

The function $\mathcal{I}_\omega$ and the matrix $\mathcal{M}_{ab}$ are defined as
\be\label{eq_M_def}
\mathcal{M}_{ab}=\sum_{c=1}^{\alpha-1} \frac{e^{-i2\pi c(a-b)/\alpha}}{\alpha\mathcal{I}_{c/\alpha}}\, \,, \hspace{4pc}
\mathcal{I}_{a/\alpha}=\int_{x_1}^{x_2}\dd x\, s_{a/\alpha}(x)\, .
\ee
Here $s_\omega(x)$ solves the following integral equation
\be
\int_0^L\dd y\,(\Phi(x,y)+\Phi_{a/\alpha}(x,y))s_{a/\alpha}(y)=\begin{cases}\,\,1 & \hspace{2pc} x_1\le x\le x_2 \\ \,\,0 & \hspace{2pc} \text{otherwise} \end{cases}\, .
\ee
The operator $\Phi$ is nothing but the connected correlation function in the ground state, defined on the interval $[0,L]$,
\be
\Phi(x,y)=\langle \hat{\phi}(x)\hat{\phi}(y)\rangle-\langle \hat{\phi}(x)\rangle\langle\hat{\phi}(y)\rangle  \,.
\ee
$\Phi_\omega$ is also defined on the interval $[0,L]$ as
\be\label{eq_def_thetaj}
\Phi_{\omega}(x,y)= \Phi(x,y) e^{i2\pi \omega (\chi_I(y)-\chi_I(x))} ,
\ee
where $\chi_I$ is the characteristic function of the interval $I$: $\chi_I(x) = 1$ if $x\in I$ and $\chi (x)= 0$ otherwise.

The last additive constant $C_\alpha$ was overlooked in Ref. \cite{bastianello2019renyi}, where the primary focus was on the coordinate dependence of the expression. The validity of the results in Ref. \cite{bastianello2019renyi} holds true apart from an overall constant, which is left ambiguous in the definition of the multi-sheet partition function defining the $\alpha-$R\'enyi entropy. However, determining the exact value of the entropy is essential for the applicability of our method (in particular for the results in \ref{sec_ent_epsilon}): the constant ambiguity can be removing simply by imposing that, in the limit $x_1\to0$, the results of Section \ref{sub_ee_edge} are recovered. This calculation is carried out in Appendix \ref{app_constant_Renyi}, resulting in
\be\label{eq_c_alpha}
C_\alpha=-\frac{1}{1-\alpha}\log\left[\frac{\alpha^{1/2}}{(2\pi)^{\alpha-1}}\right]\, .
\ee

Again, we emphasize that, in practice, we evaluate all these expressions numerically, by discretizing the system on a finite grid (Appendix \ref{app_num_discretelutt}). Thus, we are always manipulating large but finite matrices, and formula (\ref{eq_reny_int}) is evaluated through standard linear-algebra routines. 
While taking the continuum limit, the logarithm of the ratio of determinants splits into a universal part and a UV-divergent part $\propto \log \ell$, where $\ell$ is the lattice spacing. 
All the other quantities appearing in Eq. \eqref{eq_reny_int} are well defined in the continuum limit: it is possible to show \cite{bastianello2019renyi} that $s_\omega(x)$ acquires power-law singularities at the edges of the interval $[x_1,x_2]$, however these divergences remain integrable $\sim 1/|x-x_{1,2}|^{\max(\omega,1-\omega)}$ for any $0<\omega<1$. We observe that our choice of considering the $\alpha=2$ R\'enyi entropy guarantees the best convergence of the expression \eqref{eq_reny_int}.

\section{Entanglement entropy in the inhomogeneous XXZ spin chain}

Now that we have presented the general ideas and methods, we turn to a specific microscopic model where we can check our approach numerically. We focus on the XXZ spin chain in a smoothly varying external magnetic field
\be\label{eq_XXZ_ham}
\hat{H}^{\text{XXZ}}=\sum_{j=1}^{N} \left(\hat{S}^x_j\hat{S}^x_{j+1}+\hat{S}^y_j\hat{S}^y_{j+1}+\Delta \hat{S}^z_j\hat{S}^z_j \right) + \sum_{j=1}^{N} h(j/N-1/2) \hat{S}^z_j \, .
\ee
Here $\hat{S}^{x,y,z}$ are the spin-1/2 operators, and $N$ is the spin chain's length, assumed to be large. The local magnetic field $h(x)$ is some smooth function, so that $h(j/N-1/2)$ varies very slowly when $N$ is large (we shift the argument of the magnetic field in such a way the argument of $h(x)$ is zero when $j=N/2$). We stress that our chain has open boundary conditions. All numerical calculations in the chain were performed using a DMRG algorithm based on the Itensor \cite{Itensor} C++ library.

When the magnetic field $h$ is constant, the model is \emph{integrable} \cite{gaudin_2014,korepin1997quantum}: this provides a powerful set of tools which allows an exact determination of its spectrum, together with an exact computation of the Luttinger parameters $v,K$, as sketched in Appendix \ref{app_bethe ansatz}. 
Within the paramagnetic phase $|\Delta|<1$, the ground state of the homogeneous XXZ chain is gapless for $|h|<1+\Delta$. In our numerics we focus on the values $\Delta=\pm 1/2$ and $h>0$, but other choices of the parameters would lead to similar results, as long as one stays in the gapless regime. 
For details on the bosonization of the XXZ spin chain we refer to the existing literature (e.g. \cite{giamarchi2004quantum,cazalilla2004bosonizing}). Here we only need to determine the local velocity of gapless excitations $v(x)$ and the Luttinger parameter $K(x)$, which we do through the LDA.

Hereafter, we chose the magnetic field in such a way that only a subregion of the spin chain has non-trivial magnetization $-0.5<m<0.5$. The magnetization is maximal outside that region, and the spins do not fluctuate there. 
The inhomogeneous Luttinger liquid decribes only the subregion of the spin chain where the spins are critically fluctuating. Thus we stress that the spatial domain in which our Luttinger liquid lives is not the whole chain but only that subregion, which is the only one giving non-zero contribution to the entanglement entropy to the leading order (see Fig. \ref{fig_EE_UVreg} and Fig. \ref{fig_EE_cutoff}). The subleading edge behavior of the entropy corresponding to near-saturating magnetic field can also be obtained exactly \cite{stephan2019,Eisler_2014,TracyWidom_tw}, but belongs to a different edge universality class. Our focus is on bulk Luttinger liquid behavior in this paper.
In this section the bipartition consists of the two intervals $I = \{1,\dots, j \}$ and $\bar{I} = \{j+1,\dots, N \}$ of spins  in the XXZ chain (\ref{eq_XXZ_ham}).

\subsection{$\epsilon$-regularized (R\'enyi) entanglement entropy in the inhomogeneous XXZ spin chain}
\label{sec_ent_epsilon}

We approach the field theory using the following flat discretization (i.e. in the isothermal coordinates Eq. \eqref{eq_iso}),
\be\label{eq_dis_LL}
\hat{H}^{\text{LL}(\ell)}=\frac{\ell}{2\pi}\sum_{j=1}^M \pi^2 K_j\hat{\Pi}_j^2+\frac{\ell}{\pi} \sum_{j=1}^{M+1}\frac{1}{K_j+K_{j-1}}\left(\frac{\hat{\phi}_{j-1}-\hat{\phi}_{j}}{\ell}\right)^2 ,
\ee
where the discrete fields satisfy the canonical commutation relations
\be
[\hat{\phi}_j,\hat{\Pi}_{j'}]=i\delta_{j,j'},
\ee
and $\ell=M^{-1}$ is the lattice spacing is isothermal coordinates.
We impose Dirichlet boundary conditions
\be
\hat{\phi}_{0}=\hat{\phi}_{M+1}=0 .
\ee
Further details about our discretization of the free boson theory are deferred to Appendix \ref{app_num_discretelutt}.
Of course, both the XXZ spin chain and the (discretized) Luttinger Liquid bear their own cutoffs, however, in the same spirit of Eq. \eqref{eq_reg_en}, from the equation Eq. \eqref{eq_scal_ent} we can construct cut-off independent identities such as
\be\label{eq_reg_comparison}
S^{\text{XXZ}(\epsilon)}_\alpha (0,j)=S^{\text{LL}(\ell)(\epsilon)}_\alpha (0,j)
\ee
where, making use the entanglement entropy in the XXZ model $S_\alpha^\text{XXZ}$  and in the discretized Luttinger $S_\alpha^{\text{LL}(\ell)}$, we defined the regularized entropy
\be\label{eq_boundReg}
S^{(\epsilon)}_\alpha (0,j)=\frac{1}{2}\left[S_\alpha (0,j)+S_\alpha(0,j+\epsilon)-S_\alpha(j,j+\epsilon)\right]\, .
\ee
While Eq. \eqref{eq_reg_comparison} is expected to hold true for arbitrary R\'enyi entropies (and therefore also in the limit corresponding to the Von Neumann $\alpha\to 1$), computing Eq. \eqref{eq_boundReg} requires the knowledge of the entropies of an interval embedded in the bulk $(j,j+\epsilon)$. In this case, the non trivial compactification of the bosonic field plays an important role and, to the best of our knowledge, Eq. \eqref{eq_reny_int} \cite{bastianello2019renyi} is the only available result for open boundary conditions: this allows us to address the R\'enyi entropies of integer indexes, but leaves out the Von Neumann entropy from our analysis.
\begin{figure}[t!]
\includegraphics[width=0.5\textwidth]{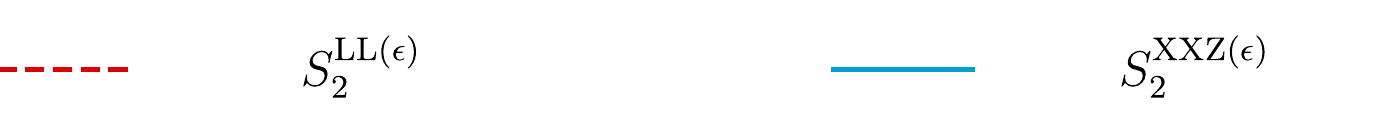}\ \\
\includegraphics[width=0.3\textwidth]{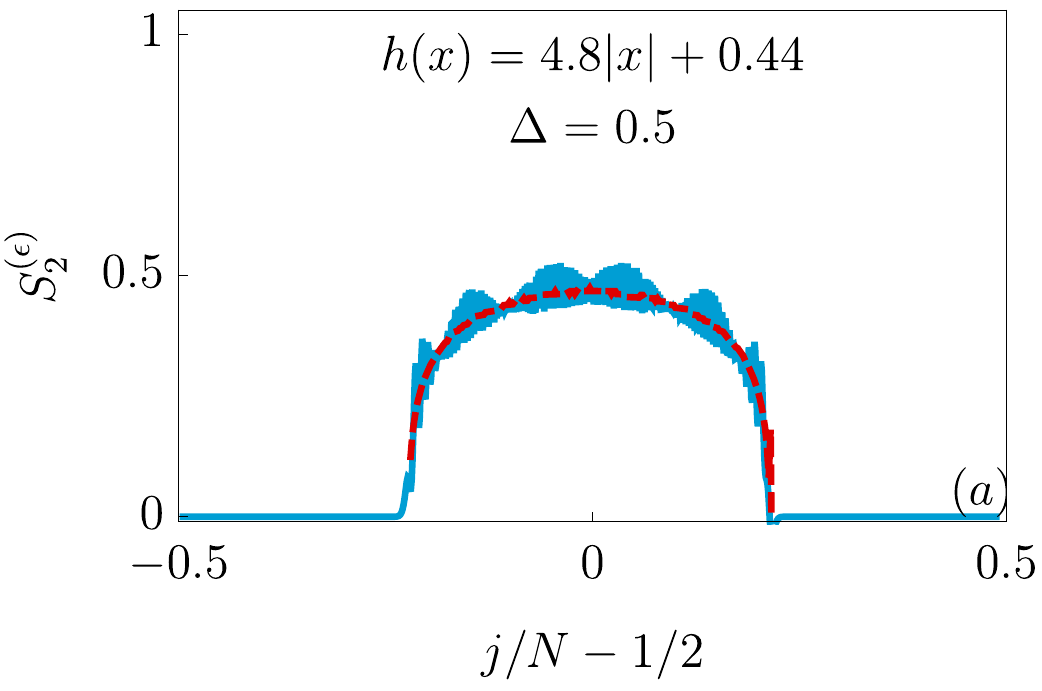}\hspace{0.5pc}
\includegraphics[width=0.3\textwidth]{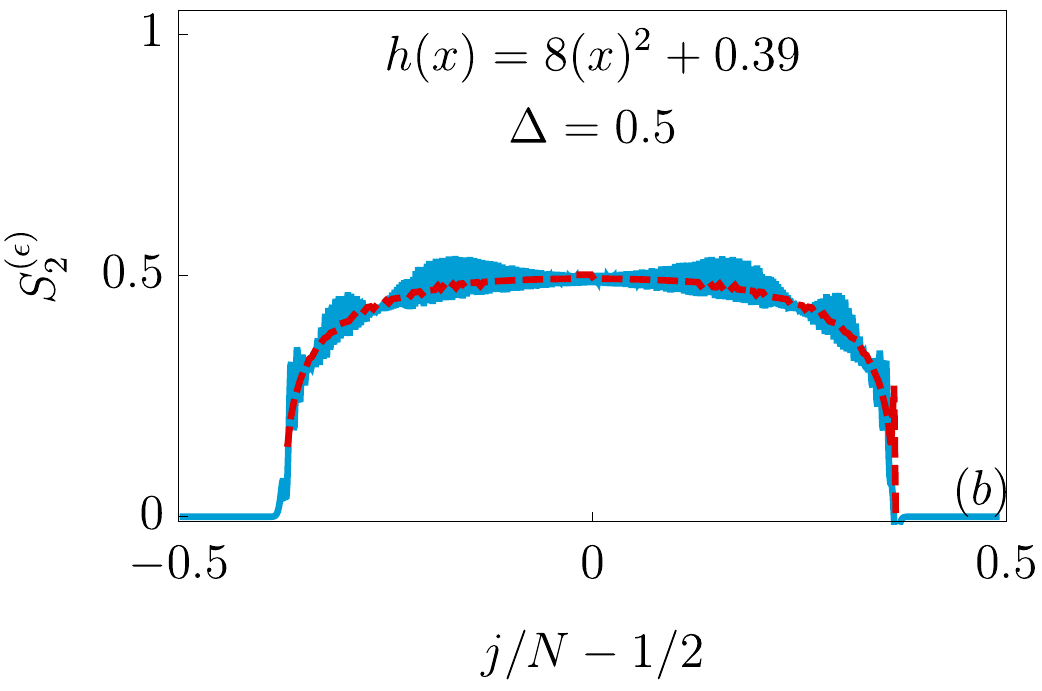}\hspace{0.5pc}
\includegraphics[width=0.3\textwidth]{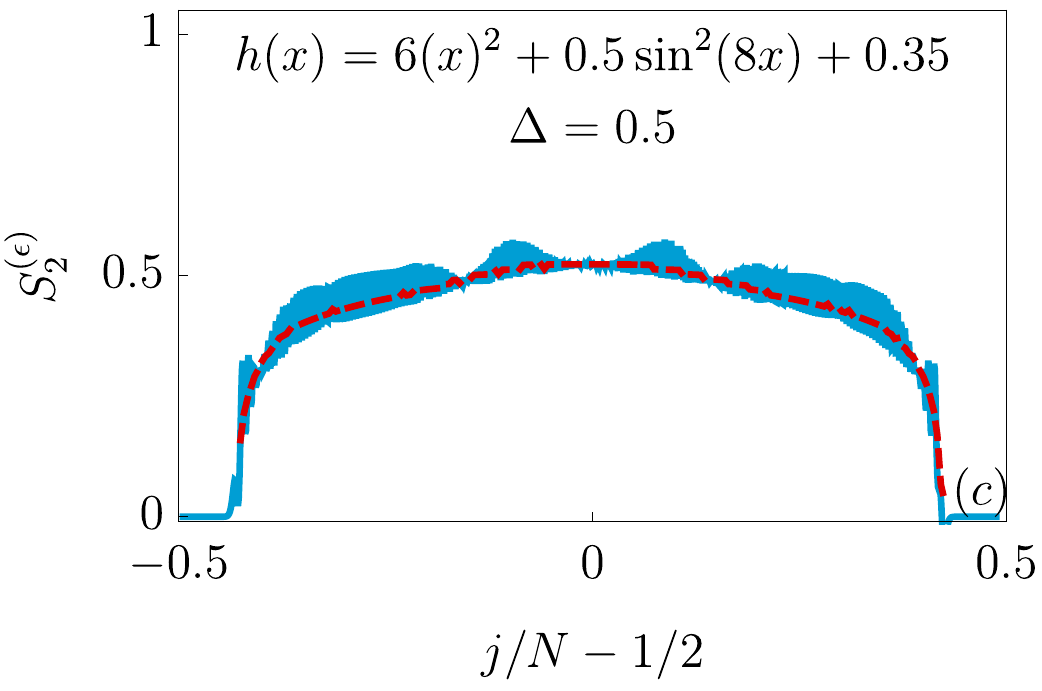}\ \\
\includegraphics[width=0.3\textwidth]{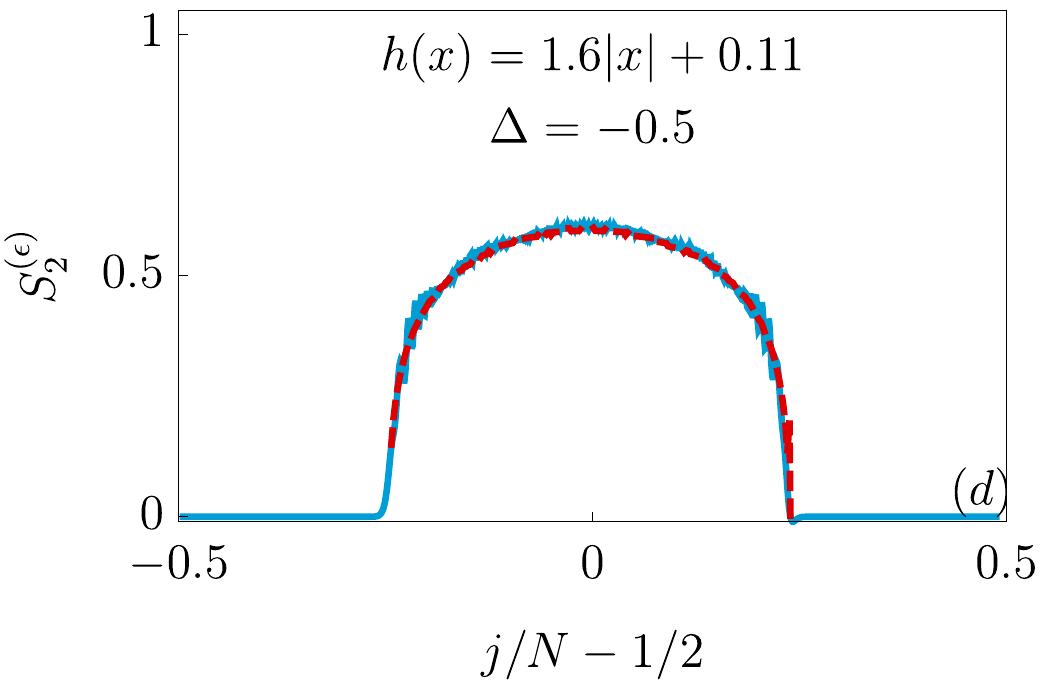}\hspace{0.5pc}
\includegraphics[width=0.3\textwidth]{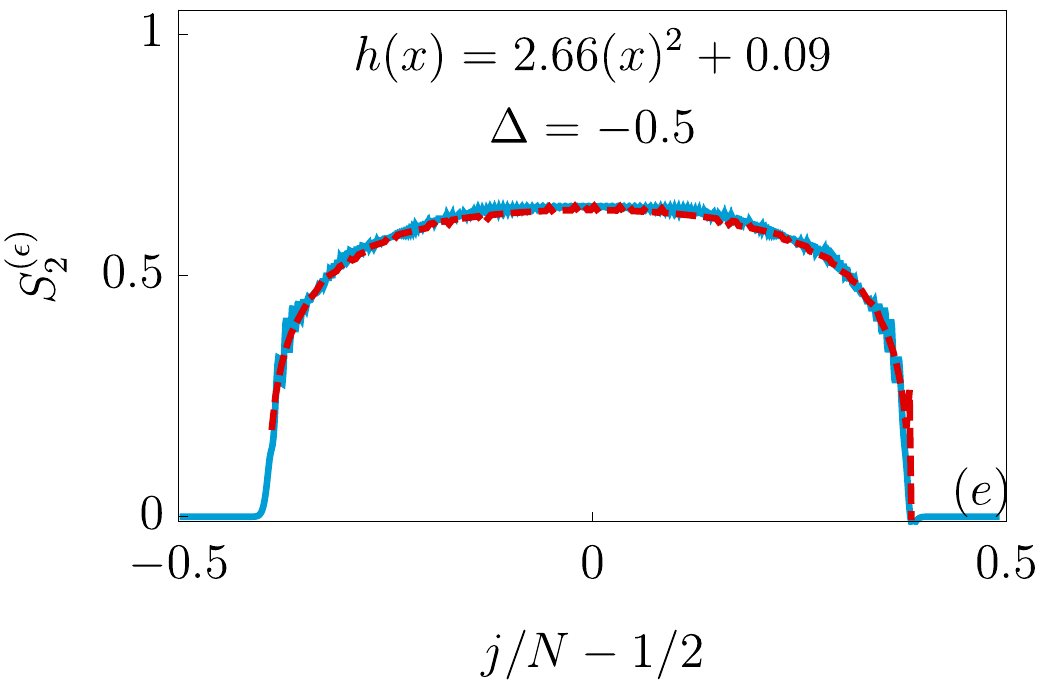}\hspace{0.5pc}
\includegraphics[width=0.3\textwidth]{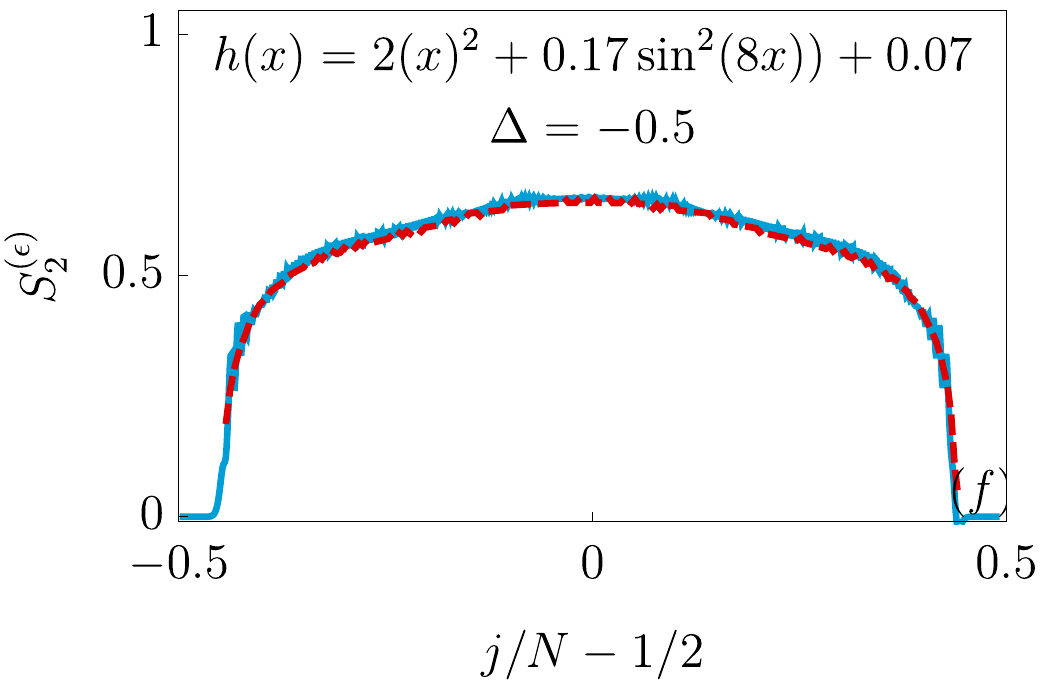}
\caption{\label{fig_EE_UVreg}Comparison of $\epsilon-$regularized entanglement entropies for the XXZ spin chain (solid blue line) and discretized Luttinger Liquid (dashed red line). We focus on the second R\'enyi entropy $S_2$, testing $\Delta=\pm 0.5$ for different magnetic traps (see inset of each plot). The comparison is excellent for $\Delta=-0.5$, while for $\Delta=0.5$ there are persistent oscillations. Indeed, as discussed in the text, corrections to the scaling limit in the form of oscillating terms are expected \cite{ee_parityXXZ} and they are enhanced as $\Delta\to 1$. The R\'enyi of the small interval $[j,j+\epsilon]$ used in the regularization suffers from the mentioned corrections, due to the small size of the interval.
The data for the XXZ spin chain are obtained for $N=1024$ sites, the small regulator is chosen homogeneous and encompassing a few lattice sites $\epsilon=8$.
In the discretized Luttinger liquid, we used isothermal coordinates and $2000$ lattice sites, resulting in $\ell=5\times 10^{-4}$: such a small discretization is required to faithfully describe the small interval of length $\epsilon$ in the original spin chain.
The small asymmetry of the curve around $j=N/2$ is induced by the $\epsilon-$regularization which explicitly breaks parity invariance.
}
\end{figure}
In Fig. \ref{fig_EE_UVreg} we compare the regularized R\'enyi entropy \eqref{eq_boundReg} for $\alpha=2$ in the XXZ model and in the discretized Luttinger, for different values of the interaction $\Delta$ and the magnetic trap, finding good agreement.
One can observe oscillations, which are much more pronounced for $\Delta=0.5$ than for $\Delta=-0.5$. This observation is consistent with known results for corrections to scaling in homogeneous systems  \cite{ee_parityXXZ}, which are controlled by the Luttinger parameter. While they disappear in the thermodynamic limit, the corresponding exponent becomes smaller as $K$ is smaller, so they significantly affect finite-size calculations when $\Delta$ approaches $1$.

\subsection{Results from numerical tabulation of the UV contribution to the entropy}
\label{sec_ent}

According to the discussion around Eq. (\ref{eq_scal_ent}) in the introduction, the entanglement entropy in the XXZ chain will be of the form
\be\label{eq_coff_obc}
S^\text{XXZ}_\alpha(0,j)=S_\alpha^\text{FT}(0,j)+c^\text{XXZ}_\alpha\big(h(j/N-1/2)\big) ,
\ee
where `FT' stands for the field theory result, which itself depends on a field theory cutoff. To evaluate the field theory part we rely on the discrete Luttinger liquid Hamiltonian (\ref{eq_dis_LL}): the entanglement entropy in the  discretized Luttinger liquid (which we note `LL$(\ell)$') satisfies a relation similar to Eq. \eqref{eq_coff_obc}, with a different non-universal $O(1)$ part $c_\alpha$,
\be
S^{\text{LL}(\ell)}_\alpha(0,j)=S_\alpha^{\text{FT}}(0,j)+c^{\text{LL}(\ell)}_\alpha\big(v(j/N), K(j/N), \ell\big) .
\ee
Here the $O(1)$ part depends on the microscopic parameters of the discretized Luttinger liquid: the lattice spacing $\ell$, and the local parameters $v(x)$ and $K(x)$ entering the Hamiltonian. Thus, we have
\be
\label{eq_res_XXZ}
S^\text{XXZ}_\alpha(0,j)=S^{\text{LL}(\ell)}_\alpha(0,j)+d_\alpha^{\text{XXZ}/{\text{LL} (\ell)}} \big(h(j/N-1/2)\big),
\ee
where $d_\alpha^{\text{XXZ}/{\text{LL} (\ell)}} (h(x))$ is the difference of the non-universal $O(1)$ parts, $c^\text{XXZ}_\alpha(h(x))-c^{\text{LL}(\ell)}_\alpha(v(x),K(x),\ell)$. Importantly, $d_\alpha^{\text{XXZ}/{\text{LL} (\ell)}} (h(x))$ should be a function of the local parameters in the model only, so we can fix it by the Local Density Approximation; we see that it depends only on the local magnetic field $h_x$. Notice that, on the discrete Luttinger liquid side, the local parameters $v(x)$ and $K(x)$ do themselves depend on $h(x)$ through the Local Density Approximation.

To evaluate $d_\alpha^{\text{XXZ}/{\text{LL} (\ell)}} (h)$, we can focus on a homogeneous spin chain. 
For a fixed magnetic field $h$, the difference
\be
	S^\text{XXZ}_\alpha(0,j) - S^{\text{LL}(\ell)}_\alpha(0,j)
\ee
is constant as a function of $j$, up to terms that go to zero in the thermodynamic limit. We tabulate the measured constants as a function of the field $h$, and this gives us access to the function $d_\alpha^{\text{XXZ}/{\text{LL} (\ell)}} (h)$, in the homogeneous spin chain.
\begin{figure}[t!]
\includegraphics[width=0.5\textwidth]{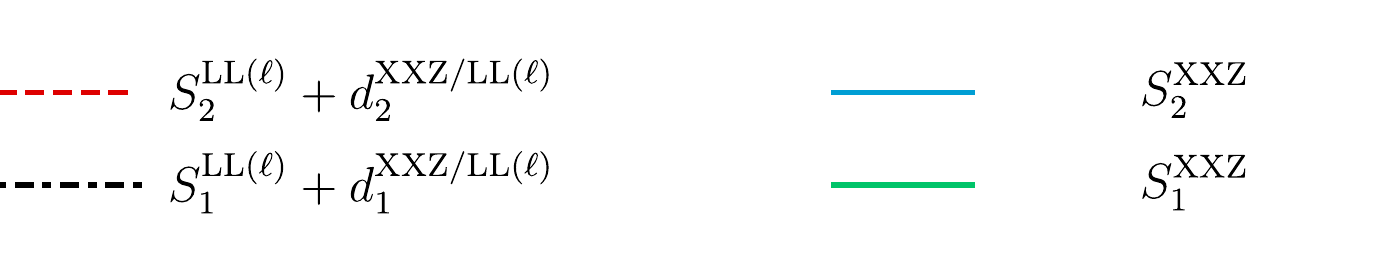}\ \\
\includegraphics[width=0.3\textwidth]{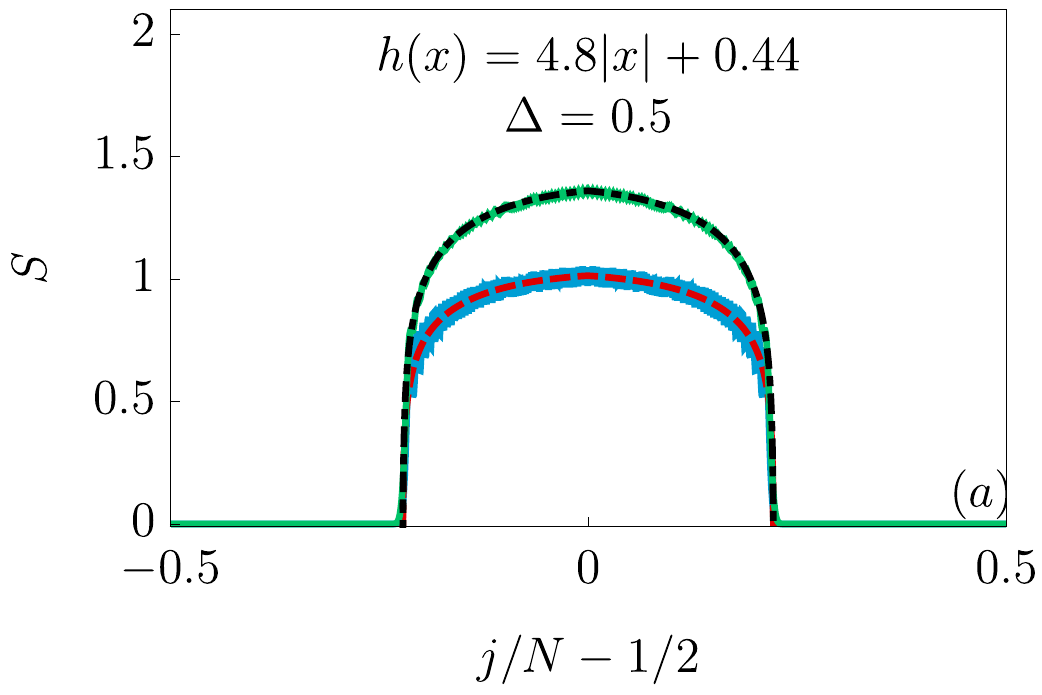}\hspace{0.5pc}
\includegraphics[width=0.3\textwidth]{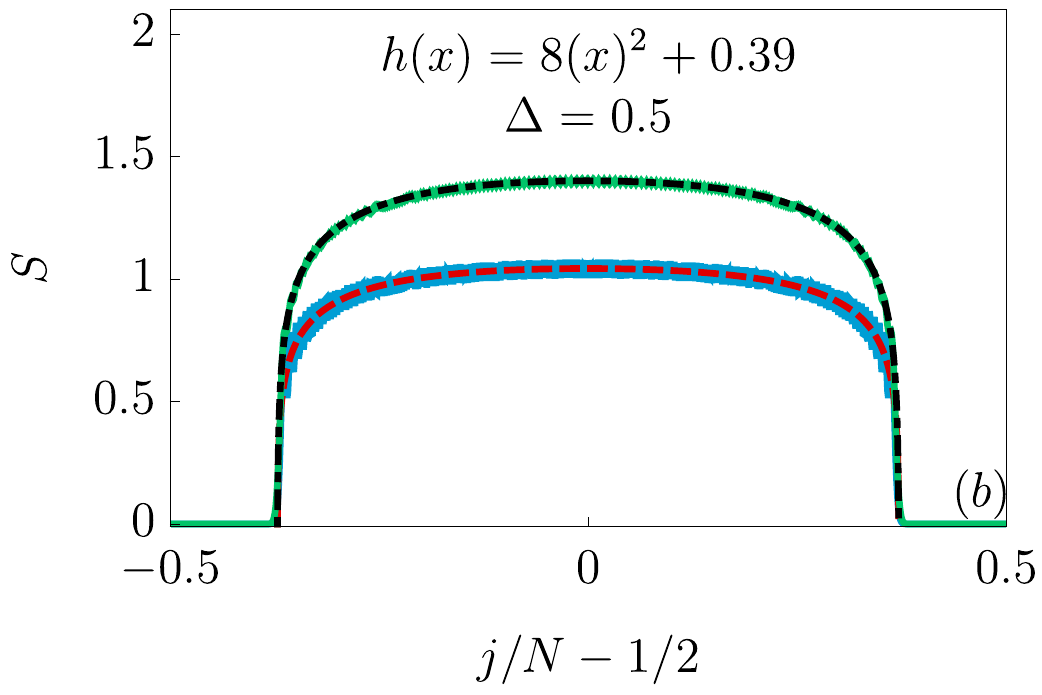}\hspace{0.5pc}
\includegraphics[width=0.3\textwidth]{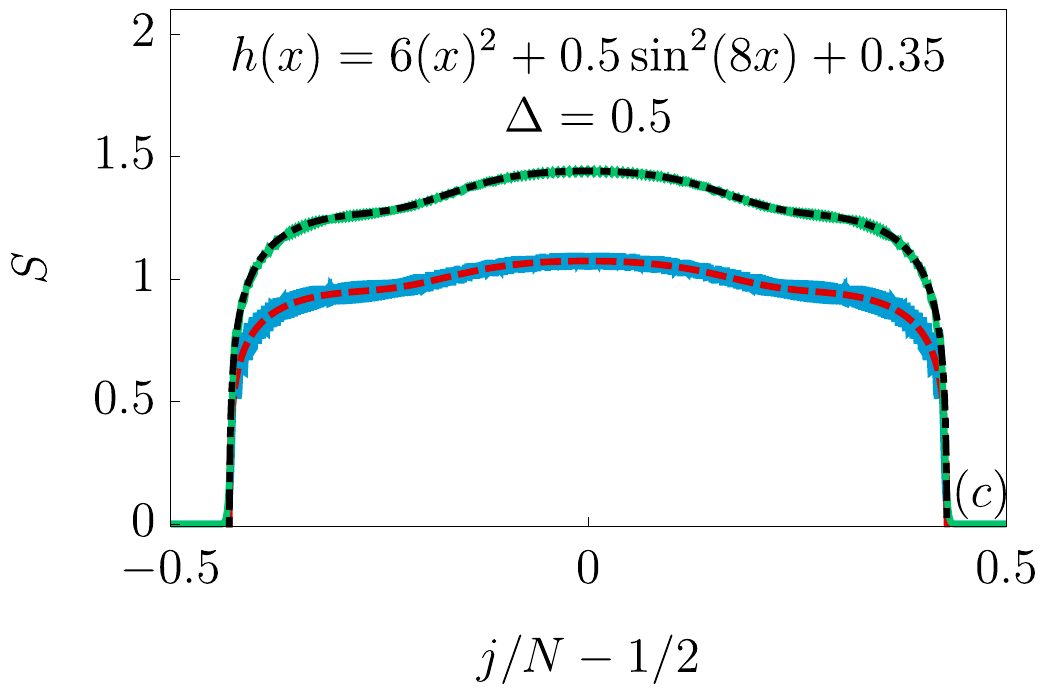}\ \\
\includegraphics[width=0.3\textwidth]{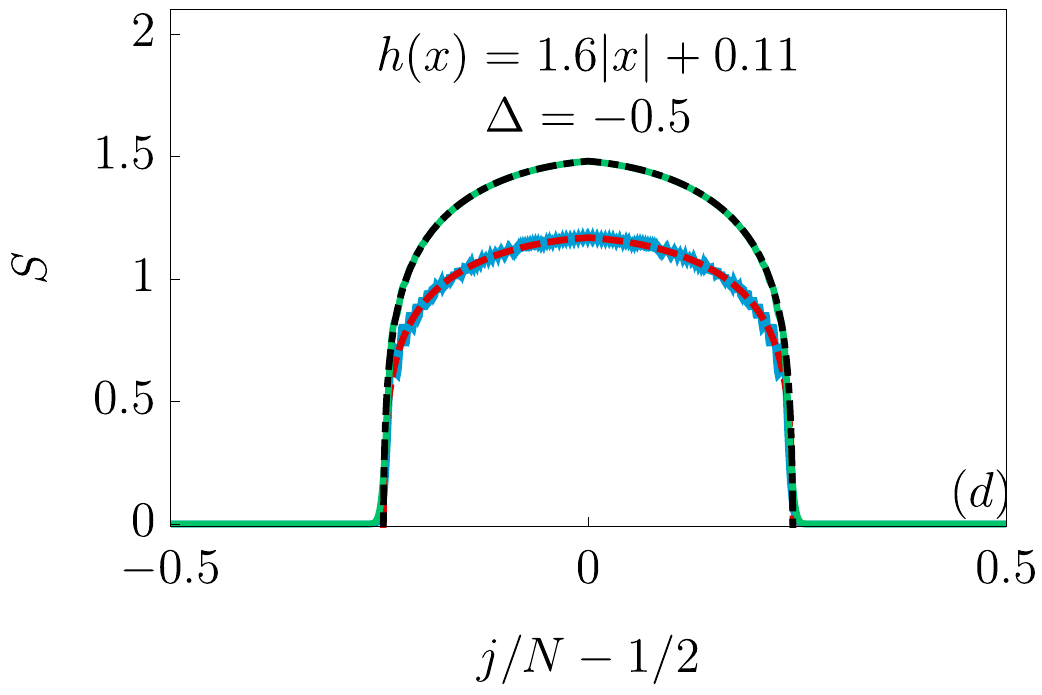}\hspace{0.5pc}
\includegraphics[width=0.3\textwidth]{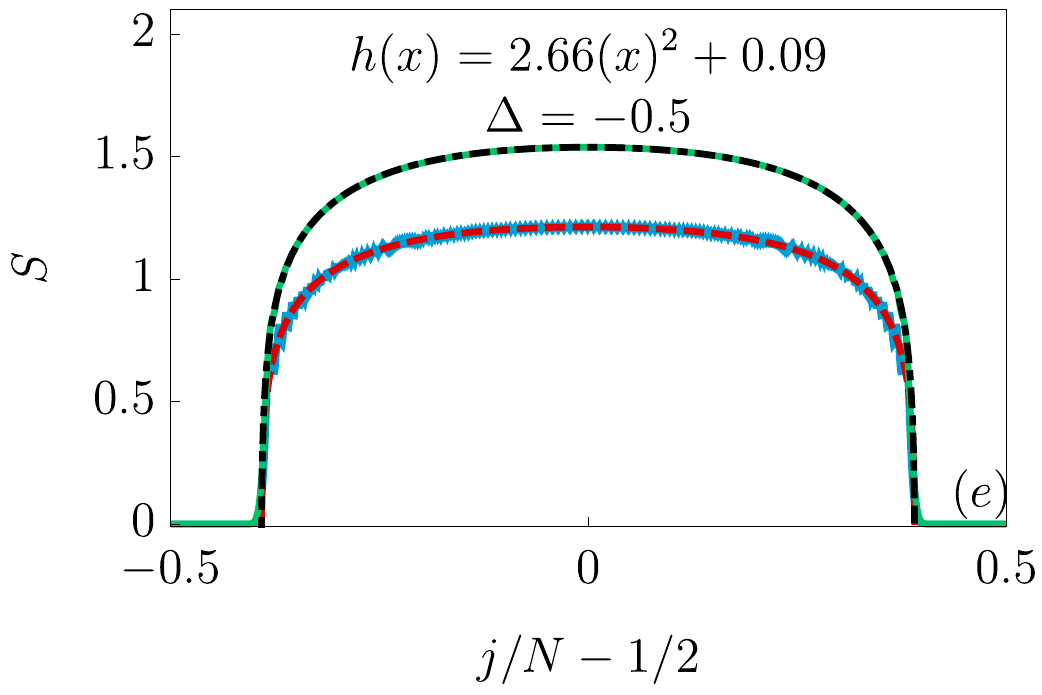}\hspace{0.5pc}
\includegraphics[width=0.3\textwidth]{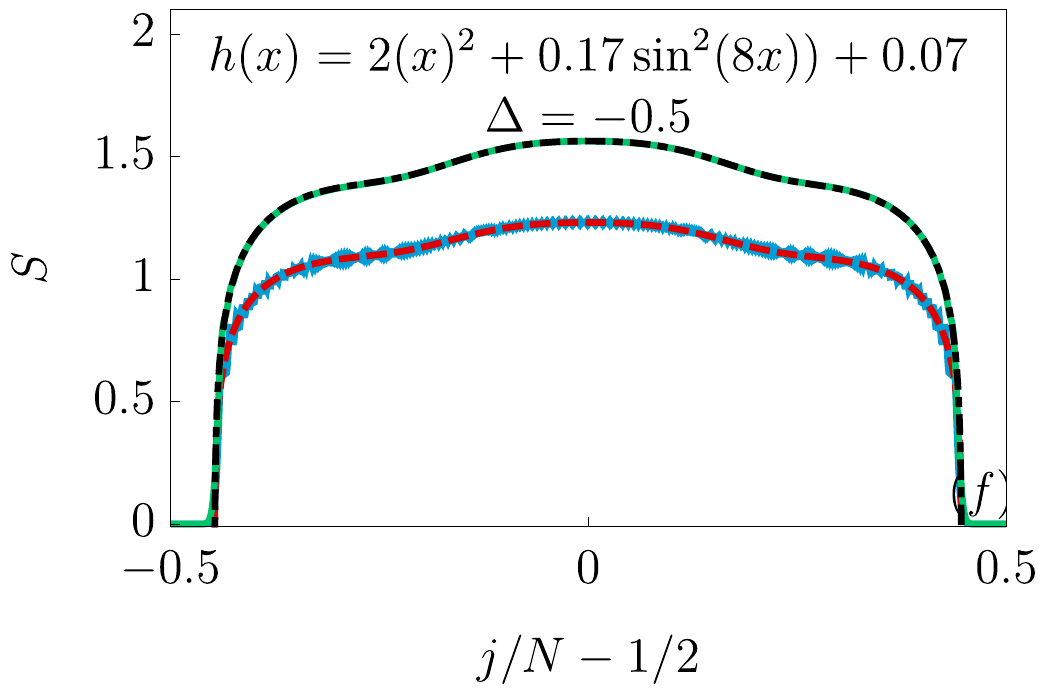}
\caption{\label{fig_EE_cutoff}
Comparison of entanglement entropies in the XXZ spin chain (solid lines) against discretized Luttinger (dashed lines) through the method of the cutoff extraction Eq. \eqref{eq_res_XXZ}. We focus on the Von Neumann entropy $S_1$ and on the second R\'enyi $S_2$ for $\Delta=\pm 0.5$ and several magnetic trap (see insets).
The XXZ spin chain has $N=1024$ lattice sites. In the case of the discretized Luttinger liquid we considered isothermal coordinates and used $200$ sites, which were enough to reach convergence. In contrast with Fig. \ref{fig_EE_UVreg}, the XXZ data show smaller oscillations around the Luttinger liquid prediction.}
\end{figure}

There is one minor subtlety that is left to discuss when one goes from the homogeneous to the inhomogeneous case. Indeed, the discrete Luttinger liquid Hamiltonian does not need to rely on the spatial discretization corresponding to the spin chain one. In general there is a change of variables $j \rightarrow \tilde{x}$ such that the spatial discretization of the discrete Luttinger liquid corresponds to $\tilde{x} \in \ell \mathbb{Z}$. Then one needs to keep track of that change of variables. This is done as follows. Recall that in field theory the calculation of entanglement entropy maps to the one of a correlation function of twist fields \cite{cardy2008form} with scaling dimension $\frac{\alpha-1/\alpha}{12}$. Then a change of variables generates an additive term in the final result for the entanglement entropy, which is the logarithm of the Jacobian of the change of variables, times the scaling dimensions. Thus, if we want to use a general coordinate $\tilde{x}$ for the discretized Luttinger liquid, Eq. (\ref{eq_res_XXZ_cov}) becomes instead
\be
\label{eq_res_XXZ_cov}
S^\text{XXZ}_\alpha(0,j)=S^{\text{LL}(\ell)}_\alpha(0,\tilde{x})+d_\alpha^{\text{XXZ}/{\text{LL} (\ell)}} (h(j/N-1/2))  + \frac{1+1/\alpha}{12}\log \left| \frac{\dd \tilde{x}}{ \dd j} \right|.
\ee

In Fig. \ref{fig_EE_cutoff} we compare the data of several DMRG simulations of the full inhomogeneous system, with those extracted from the (discretized) Luttinger liquid, using Eq. (\ref{eq_res_XXZ_cov}). We focus on the Von Neumann entanglement entropy $S_1$ and the R\'enyi entropy $S_2$: the agreement is excellent in both cases.

\section{Conclusion}
\label{sec_concl}

We have developed a method to calculate the entanglement entropy of Luttinger liquids in inhomogeneous situations, where the Luttinger parameter $K$ becomes position dependent. The crucial ingredient in our analysis is a careful treatment of the contribution to the entanglement entropy coming from the UV cutoff, both in the microscopic model one is interested in (in this paper, the XXZ chain) and in the discretized free boson field theory one uses to do the calculations. Contrary to homogeneous situations, where this contribution is simply a constant often treated as a fitting parameter, here that constant depends on position. It is possible to construct combinations of entanglement entropies, like the $\epsilon$-regularized entropy we considered (see also Refs.~\cite{bianchi2014entanglement,bianchi2015entanglement} for similar definitions of regularized entanglement entropies), such that the non-universal position-dependent constant disappears. Alternatively, we showed that this position-dependent contribution to the entanglement entropy can be tabulated, and then used in numerics to arrive at concrete predictions for the standard (i.e. not regularized) entanglement entropy.
In addition, our analysis provides a numerical benchmark for the results of Ref. \cite{bastianello2019renyi} in the interacting case ($K\ne 1$), which was not tested in the original reference.

It would be very interesting to extend our results to deal with out-of-equilibrium inhomogeneous situations in critical spin chains or one-dimensional quantum gases. One paradigmatic example of such situations corresponds to the Riemann problem in hydrodynamics, where two semi-infinite systems with different thermodynamic characteristics are suddenly joined together \cite{antal1999transport,antal2008,platini2007relaxation,bertini2016transport,piroli2017transport,collura2018analytic}. In the particular case of two initial half-systems at zero temperature (but, say, at different filling fractions), the system must be described by a time-dependent inhomogeneous Luttinger liquid. The free fermion case is mostly understood \cite{dubail2017conformal}, however in the truly interacting case
where the Luttinger parameter $K$ is expected to depend on $x$ and $t$, there is no theory allowing a calculation of the entanglement entropy, to our knowledge. A number of numerical observations on entanglement entropies have been made \cite{sabetta2013nonequilibrium,alba2014entanglement,gruber2019magnetization,eisler2019front}, and it would be very nice to adapt our results to the time-dependent case and make contact with those works. Here again, the non-universal position-dependent (and also time-dependent) contribution to the entanglement entropy will require a careful treatment. This time, to tabulate it one might need to look at homogeneous states that are not just the ground state of the homogeneous problem, but highly excited states. In particular, in some cases it might be necessary to go beyond the simple Luttinger liquid and consider so-called ``split Fermi seas'' \cite{fokkema2014split,vlijm2016correlations,eliens2016general,eliens2017quantum,QGHD}, and calculate entanglement entropies in those, which could be hard. Alternatively, focusing on $\epsilon$-regularized entanglement entropies should be easier in those out-of-equilibrium situations; this could be a good starting point to investigate this class of problems.

\vspace{1cm}

{\bf Acknowledgments.} We are grateful to Paola Ruggiero, Jacopo Viti and Pasquale Calabrese for inspiring discussions. We are especially grateful to Paola Ruggiero for pointing out Refs.~\cite{bianchi2014entanglement,bianchi2015entanglement}.
A.B. acknowledges the support from the European Research Council under ERC Advanced grant 743032 DYNAMINT.

\appendix

\section{Inhomogeneous Luttinger liquid: from Lagrangian to Hamiltonian}
\label{app:S_H}

In this appendix we work with Lorentzian signature. The action of the inhomogeneous Luttinger liquid is
\begin{eqnarray}
\nonumber	S &=& \frac{1}{2 \pi} \int \frac{\dd x \dd t}{K} \sqrt{- g} \, g^{ab} (\partial_a \phi ) (\partial_b \phi) \\
	 &=& \frac{1}{2 \pi} \int \frac{\dd x \dd t}{K} \sqrt{- g} \, [ g^{xx} (\partial_x \phi )^2+ g^{tt} (\partial_t \phi )^2 + 2g^{tx} (\partial_t \phi ) (\partial_x \phi) ] ,
\end{eqnarray}
which gives the Lagrangian $L = \frac{1}{2\pi} \int dx \sqrt{- g} \, [ g^{xx} (\partial_x \phi )^2+ g^{tt} (\partial_t \phi )^2 + 2g^{tx} (\partial_t \phi ) (\partial_x \phi) ]$. The entries of the metric $g$ and the Luttinger parameter $K$ depend on $x$ and $t$. 
The canonical momentum associated to $\phi(x)$ is
$$
\Pi(x) = \frac{\partial L}{\partial ( \partial_t \phi(x))}  =  \frac{1}{ \pi K} \sqrt{- g} \, [ g^{tt} (\partial_t \phi ) + g^{tx} (\partial_x \phi) ] ,
$$
so 
$$
 g^{tt} (\partial_t \phi )   =  \frac{ \pi K}{ \sqrt{- g}} \Pi(x) -  g^{tx} (\partial_x \phi) 
$$
and the Hamiltonian derived from the action $S$ is then
\begin{eqnarray*}
H &=& \int \dd x \, \Pi \partial_t \phi -  \frac{1}{2 \pi} \int \frac{\dd x}{K} \sqrt{- g} \, [ g^{xx} (\partial_x \phi )^2+ g^{tt} (\partial_t \phi )^2 + 2g^{tx} (\partial_t \phi ) (\partial_x \phi) ]  \\
&=& \int \frac{\dd x}{\pi K}  \sqrt{- g} \, [ g^{tt} (\partial_t \phi )^2 + g^{tx} (\partial_x \phi) (\partial_t \phi) ]  -  \frac{1}{2 \pi} \int \frac{\dd x}{K} \sqrt{- g} \, [ g^{xx} (\partial_x \phi )^2+ g^{tt} (\partial_t \phi )^2 + 2g^{tx} (\partial_t \phi ) (\partial_x \phi) ]  \\
&=& \int \frac{\dd x}{2\pi K}  \sqrt{- g} \, [ g^{tt} (\partial_t \phi )^2  -  g^{xx} (\partial_x \phi )^2  ]  \\
&=& \int \frac{\dd x}{2\pi K}  \sqrt{- g} \, \left[ \frac{1}{g^{tt}} \left(  \frac{ \pi K}{ \sqrt{- g}} \Pi(x) -  g^{tx} (\partial_x \phi)  \right)^2  -  g^{xx} (\partial_x \phi )^2  \right]  .
\end{eqnarray*}
In the $(t,x)$ basis, the metric is the $2 \times 2$ matrix
$$
g = \left( \begin{array}{cc}
	v - \frac{u^2}{v} &  \frac{u}{v} \\
	 \frac{u}{v} & -\frac{1}{v}
\end{array} \right)
$$
where $u \pm v$ are the velocities of sound waves in the fluid. In this paper we are working with a time-independent Luttinger liquid with $u=0$; but more generally it is useful to consider the case where $u$ does not necessarily vanish. The above metric has determinant $-1$, so $\sqrt{- g} = 1$, $g^{tt} =  \frac{1}{v}$, $g^{x t} =  \frac{u}{v}$, $g^{xx} =  \frac{u^2}{v}-v$. Then the final form of the Hamiltonian is
\begin{equation}
H \, = \, \int \frac{\dd x}{2\pi K}  \left[ v \left(  -\pi K \,\hat{\Pi}(x) + \frac{u}{v} (\partial_x \hat{\phi})  \right)^2  -  (\frac{u^2}{v} - v) (\partial_x \hat{\phi} )^2  \right]  ,
\end{equation}
where $\hat{\phi}$ and $\hat{\Pi}$ are canonically conjugated: $[\hat{\phi}(x), \hat{\Pi}(y) ] \, = \, i \delta(x-y)$. When $u=0$, this is the Hamiltonian (\ref{H_inh}) in the main text.

\section{Luttinger parameters $v,K$ in the XXZ spin chain}
\label{app_bethe ansatz}

The Luttinger description captures the low energy sector of interacting one-dimensional systems, but the difficult task of determining the effective parameters $v$ and $K$ is left open. In general, one is forced to use some approximations or numerical analysis, except for an important set of models, namely integrable systems \cite{gaudin_2014,korepin1997quantum}.
Their thermodynamics can be exactly computed in terms of proper integral equations, thanks to the Thermodynamic Bethe ansatz: then, the exact solution is matched with the Luttinger prediction, providing the expressions of $v$ and $K$.
In the spirit of the local density approximation, we can compute the local $v(x)$ and $K(x)$ as if the model was homogeneous. 
Since in this work we tested our prediction on the XXZ spin chain, we restrict the discussion of the TBA and the subsequent extraction of $v$ and $K$ to this model. We closely follow Ref. \cite{franchini2017introduction}.

We take for granted a basic knowledge of TBA, for which the reader can refer to \cite{gaudin_2014,korepin1997quantum}, and just give the main formulas.
Thus, we consider the Hamiltonian \eqref{eq_XXZ_ham} for an homogeneous magnetic field $h$.

Notice that with a rotation along the $x$ axis one can send $\hat{S}^x\to \hat{S}^x$, $\hat{S}^y\to -\hat{S}^y$ and $\hat{S}^z\to-\hat{S}^z$: thus we can change the sign $h\to -h$. Therefore, we limit ourselves to study the case $h>0$ and focus on $|\Delta|<1$, which we parametrize as $\Delta=\cos(\pi \gamma)$.
Due to integrability, the model can be understood in terms of stable quasiparticle excitations which are distributed accordingly to functions known as root densities. The description of the ground state requires a single root density, which satisfies the following linear integral equation
\be\label{eq_root_def}
\rho(\lambda)=a(\lambda)+\int_{-\Lambda}^\Lambda \frac{\dd \mu}{2\pi}\, \varphi(\lambda-\mu) \rho(\mu) 
\ee
where
\be
a(\lambda)=\frac{1}{\pi}\frac{\sin(\pi\gamma)}{\cosh(2\lambda)-\cos(\pi\gamma)}\, \hspace{2pc}
\varphi(\lambda)=-2\frac{\sin(2\pi\gamma)}{\cosh(2\lambda)-\cos(2\pi\gamma)}\, .
\ee
The Fermi sea $\Lambda$ is identified by the constraint that the dressed energy
\be\label{eq_B3}
\varepsilon(\lambda)= E(\lambda)+h+\int_{-\Lambda}^\Lambda \frac{\dd \mu}{2\pi} \varphi(\lambda-\mu)\varepsilon(\mu)
\ee
is equal to zero at the Fermi point $\varepsilon(\Lambda)=0$. Above, the bare energy is $E(\lambda)=-\pi\sin(\pi\gamma) a(\lambda)$.

We now consider the problem of fixing the Luttinger parameters: the velocity of sound appearing in the Luttinger is nothing else than the dressed velocity of the excitations at the edge of the fermi sea, namely
\be
v=\frac{(\partial_\lambda E(\lambda))^\text{dr}}{(\partial_\lambda p(\lambda))^\text{dr}}\Big|_{\lambda=\Lambda}
\ee
where, for an arbitrary test function $\tau$, the dressing operation is defined through the following integral equations
\be
\tau^\text{dr}(\lambda)=\tau(\lambda)+\int_{-\Lambda}^\Lambda \frac{\dd\mu}{2\pi}\varphi(\lambda-\mu)\tau^\text{dr}(\mu)\, .
\ee
Above, we need to use that $\partial_\lambda p(\lambda)=2\pi a(\lambda)$.
In order to find $K$, we proceed as it follows.

Conveniently, the Hamiltonian \eqref{eq_XXZ_ham} can be regarded as a fermionic system with the correspondence $\hat{S}^z\to \frac{1}{2}-\hat{c}^\dagger_j\hat{c}_j$, being $\hat{c}_j$ standard spinless fermions \cite{korepin1997quantum}.  Let $n=\langle \hat{c}^\dagger_j\hat{c}_j\rangle$ be the density of Fermions, thus from Bethe Ansatz we have
\be
n=\int_{-\Lambda}^\Lambda \dd \lambda\, \rho(\lambda)
\ee
We define the compressibility as the system's response to external variations of the magnetic field $\kappa=-\partial_h n$. In the fermionic language, adding a small magnetic field $\delta h$ to the Hamiltonian couples to the fermion density, thus in the Luttinger liquid one gets a shift proportional to $\partial_x\hat{\phi}$
\be
\hat{H}\to\hat{H}'=\frac{1}{2\pi}\int_0^L \dd x \, v\,\left[  K  (\pi \hat{\Pi})^2+\frac{1}{K} (\partial_x\hat{\phi})^2\right]+\int_0^L \dd x\, \frac{\delta h}{\pi}\partial_x\hat{\phi}\, .
\ee

One can then redefine $\partial_x\hat{\phi}'=\partial_x\hat{\phi}+\frac{K}{v}\delta h$ and $\hat{H}'$ is back to the Hamiltonian for $h=0$, but expressed in terms of the new variables, thus $\langle \partial_x\hat{\phi}'\rangle_{\delta h}=\langle \partial_x\hat{\phi}\rangle_{\delta h=0}$. Therefore
\be
n(\delta h)-n(\delta h=0)=-\frac{1}{\pi}\left(\langle \partial_x\hat{\phi}\rangle_{\delta h}-\langle \partial_x\hat{\phi}\rangle_{\delta h=0}\right)=-\frac{K}{v\pi }\delta h
\ee
Thus, within the Luttinger Liquid theory we get
\be\label{eq_com_lutt}
\kappa=\frac{K}{v\pi}\, .
\ee

We now compute the same object in the Bethe Ansatz formalism. Firstly, we need to study  how the Fermi sea $\Lambda$ is affected by a change in $h$. Thus, we regard the effective energy $\varepsilon$ as an independent function of the rapidity, the magnatic field and $\Lambda$, i.e. $\varepsilon(\lambda,\Lambda,h)$. Considering a small variation in the magnetic field $h\to h+\delta h$ and in the Fermi sea $\Lambda\to\Lambda+\delta \Lambda$, we get
\be\label{eq_deltaE}
\varepsilon(\lambda,\Lambda+\delta\Lambda,h+\delta h)=\varepsilon(\lambda,\Lambda,h)+\delta h Z(\lambda,\Lambda)\, .
\ee
Above, we defined 
\be\label{z_def}
Z(\lambda,\Lambda)=1+\int_{-\Lambda}^\Lambda \frac{\dd\mu}{2\pi} \varphi(\lambda-\mu)Z(\lambda,\Lambda)
\ee
and we explicitly used $\varepsilon(\Lambda,\Lambda,h)=0$. Let us now compute Eq. \eqref{eq_deltaE} at $\lambda=\Lambda+\delta \Lambda$ imposing $\varepsilon(\Lambda+\delta\Lambda,\Lambda+\delta\Lambda,h+\delta h)=0$. One gets
\be
\delta \Lambda\partial_\lambda\varepsilon\big|_{\lambda=\Lambda}+\delta h Z(\Lambda)=0\, ,
\ee
where we neglect to write explicitly the $\Lambda,h$ dependence, since there is no risk of ambiguities.
From the above it follows
\be
\delta \Lambda=-\delta h \frac{Z(\Lambda)}{\partial_\lambda\varepsilon\big|_{\lambda=\Lambda}}\, .
\ee
We can now take the variation of the integral equation defining the root density \eqref{eq_root_def}
\be
\delta\rho(\lambda)=\delta \Lambda\left[\frac{1}{2\pi}\, \varphi(\lambda-\Lambda) \rho(\Lambda)+\frac{1}{2\pi}\, \varphi(\lambda+\Lambda) \rho(-\Lambda)\right]+\int_{-\Lambda}^\Lambda \frac{\dd \mu}{2\pi}\, \varphi(\lambda-\mu) \delta\rho(\mu) \, .
\ee

It is convenient to define the function $F(\lambda)$, which satisfies the following integral equation (we explicitly use that $\rho(\Lambda)=\rho(-\Lambda)$)
\be
F(\lambda)=\left[\frac{1}{2\pi}\, \varphi(\lambda-\Lambda) +\frac{1}{2\pi}\, \varphi(\lambda+\Lambda) \right]+\int_{-\Lambda}^\Lambda \frac{\dd \mu}{2\pi}\, \varphi(\lambda-\mu) F(\mu)\, .
\ee
In this way $\delta\rho(\lambda)=\delta \Lambda \rho(\Lambda)F(\lambda)$. Finally, we compute the variation of the density
\be
\delta n=\delta \Lambda\Big[2\rho(\Lambda)\Big]+\delta \Lambda \rho(\Lambda) \int_{-\Lambda}^\Lambda \dd \lambda \, F(\lambda)\, .
\ee
Using the integral equation it is rather simple to show
\be
\int_{-\Lambda}^\Lambda \dd \lambda \, F(\lambda)=\int_{-\Lambda}^\Lambda \dd \lambda \, Z(\lambda)\left[\frac{1}{2\pi}\, \varphi(\lambda-\Lambda) +\frac{1}{2\pi}\, \varphi(\lambda+\Lambda) \right]\, .
\ee
\begin{figure}[t!]
\includegraphics[width=0.3\textwidth]{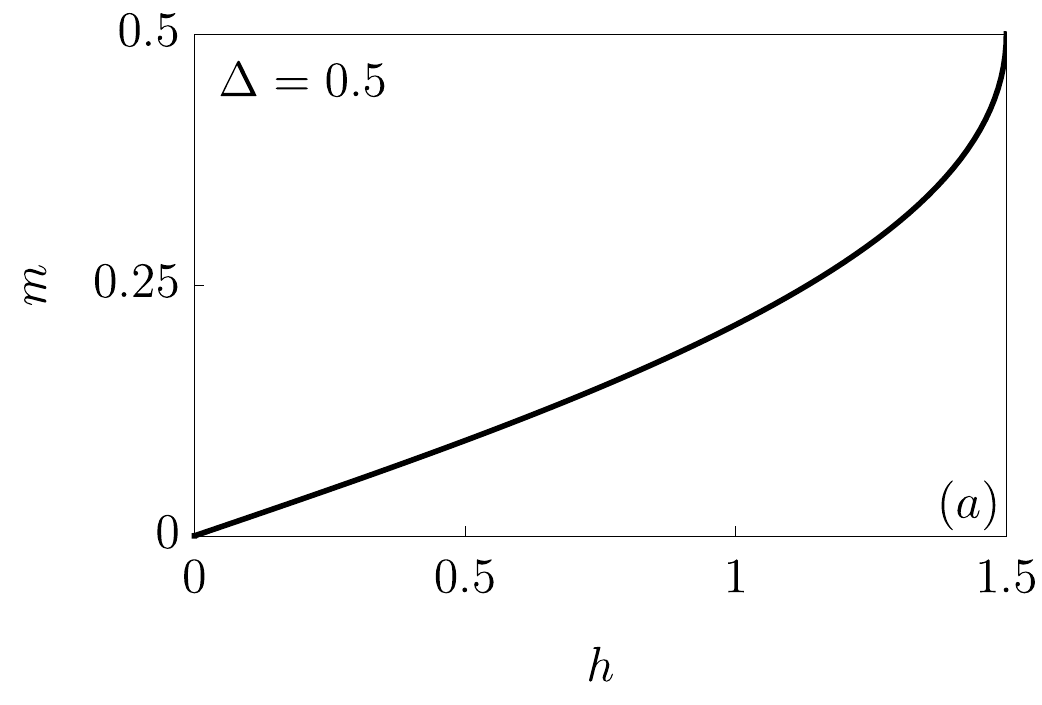}\hspace{0.5pc}
\includegraphics[width=0.3\textwidth]{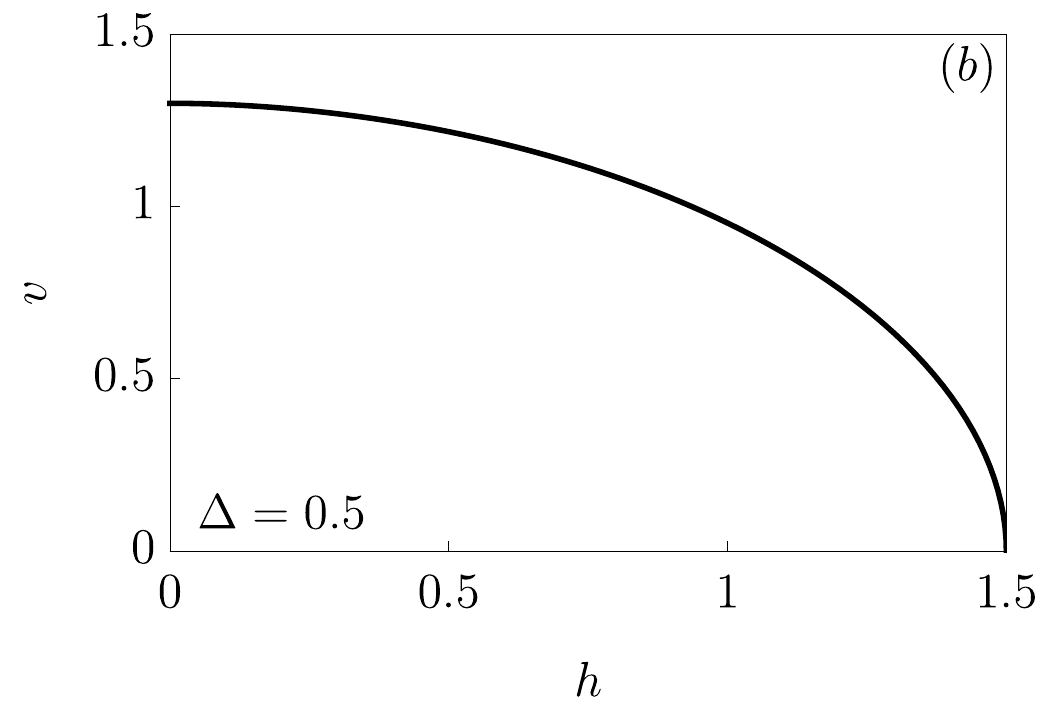}\hspace{0.5pc}
\includegraphics[width=0.3\textwidth]{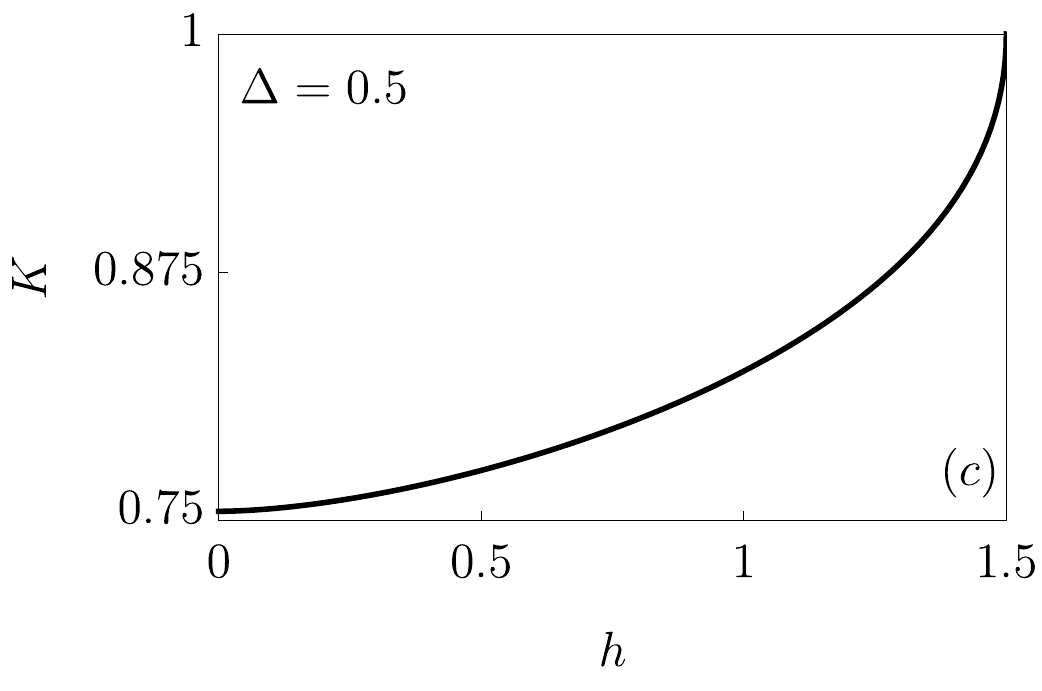}\ \\
\includegraphics[width=0.3\textwidth]{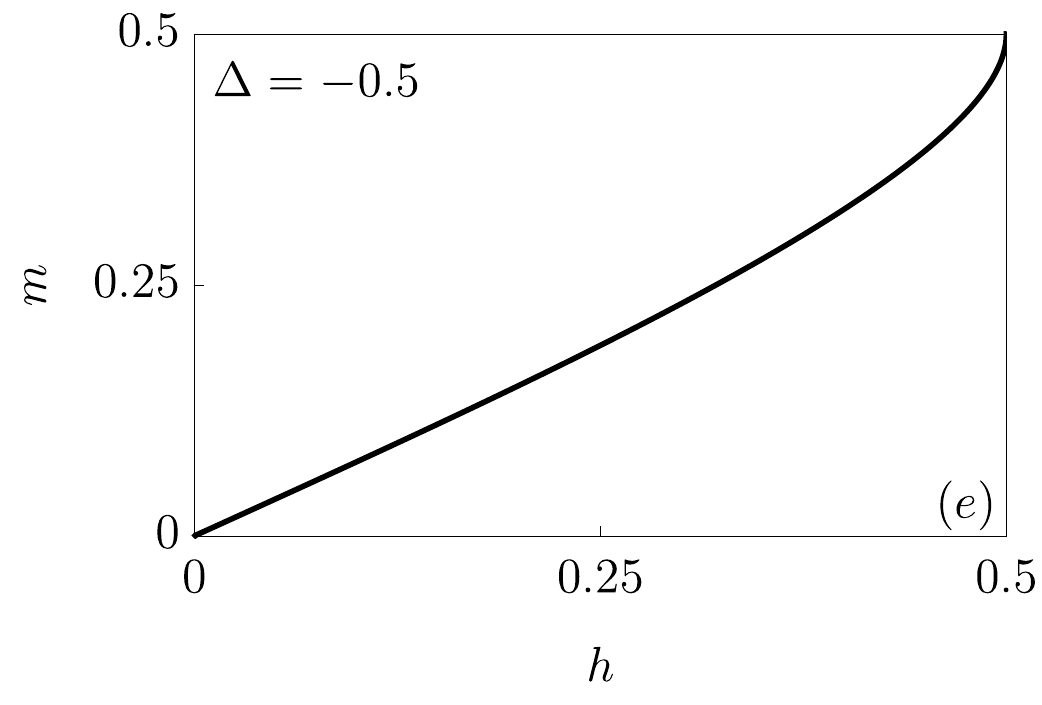}\hspace{0.5pc}
\includegraphics[width=0.3\textwidth]{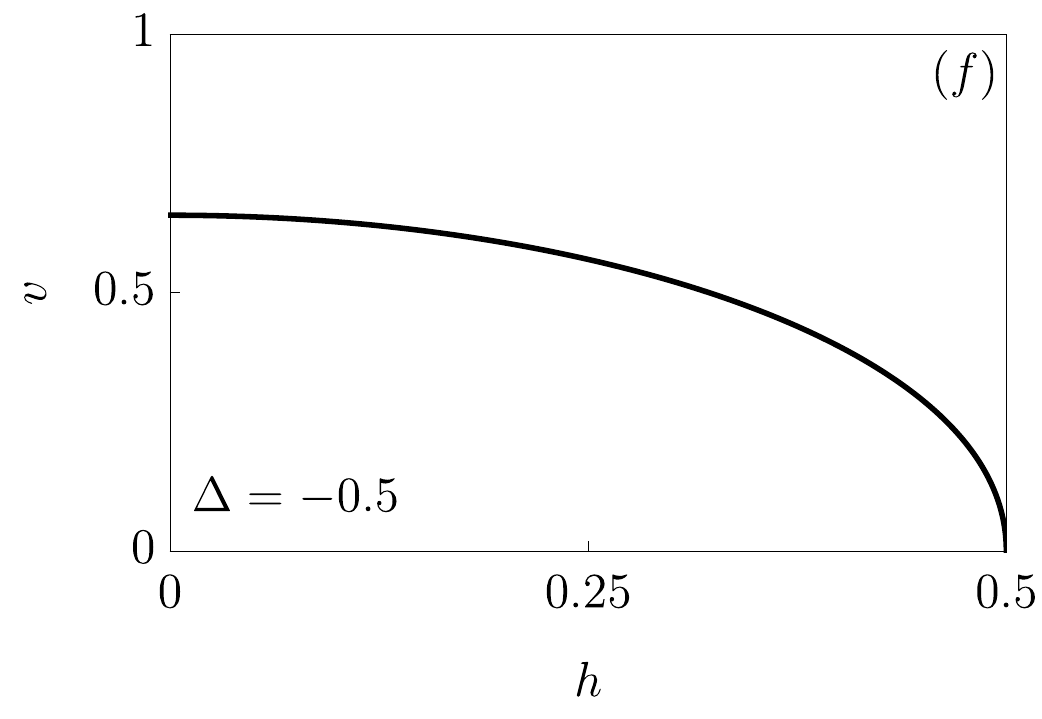}\hspace{0.5pc}
\includegraphics[width=0.3\textwidth]{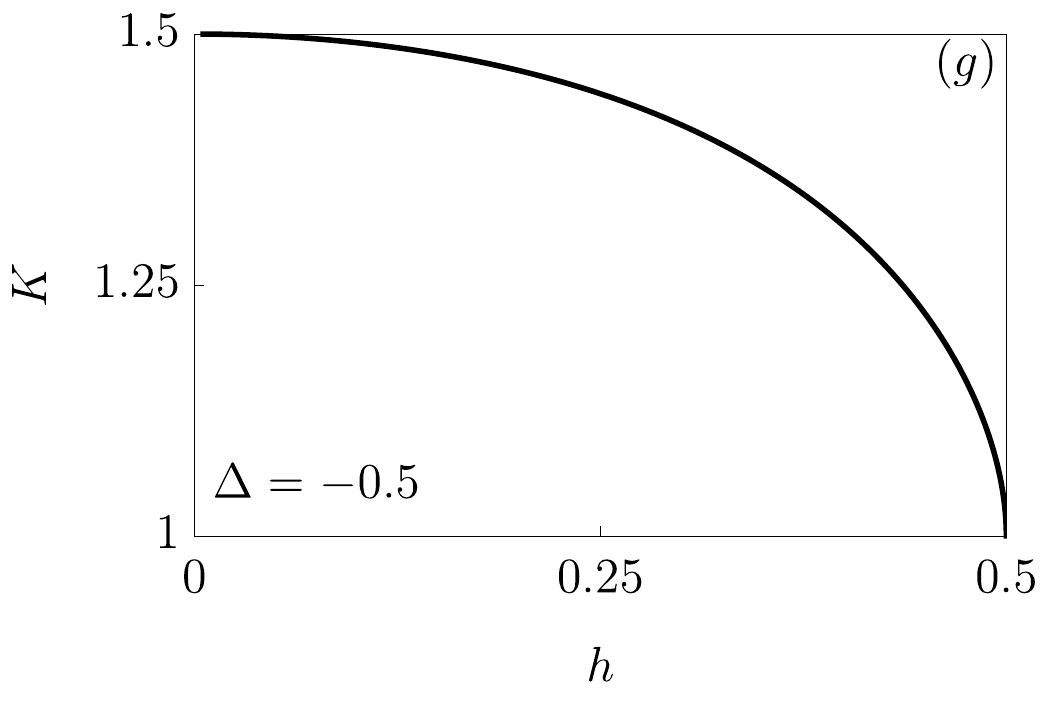}
\caption{\label{fig_LDA_par}{Luttinger parameters and local magnetization in the XXZ spin chain obtained from the Bethe Ansatz solution as a function of the local magnetic field $h$. We focus on the cases $\Delta=\pm 0.5$.
See also Ref. \cite{eisler2017front}, where similar plots are shown for different interactions.
}}
\end{figure}
Above, using the definition of $Z(\lambda)$ \eqref{z_def} and the symmetry of the kernel, one gets
\be
\int_{-\Lambda}^\Lambda \dd \lambda\,  F(\lambda)=Z(\Lambda)+Z(-\Lambda)-2\, ,
\ee
thus
\be
\delta n=\delta \Lambda \rho(\Lambda)2Z(\Lambda)\,,
\ee
where we also used $Z(\lambda)=Z(-\lambda)$. One then reaches the following expression for the compressibility
\be
\kappa=\frac{2\rho(\Lambda)Z^2(\Lambda)}{\partial_\lambda \varepsilon\big|_{\lambda=\Lambda}}\, ,
\ee
which can be further simplified. Deriving the integral equation Eq. \eqref{eq_B3}, using the symmetry of the kernel, integrating by parts and using that $\varepsilon(\pm\Lambda)=0$ one gets
\be
\partial_\lambda \varepsilon(\lambda)=\partial_\lambda E+\int_{-\Lambda}^\Lambda \frac{\dd \mu}{2\pi} \varphi(\lambda-\mu)\partial_\mu \varepsilon(\mu)\, ,
\ee
from which we can conclude $\partial_\lambda \varepsilon(\lambda)=(\partial_\lambda E(\lambda))^\text{dr}$. Using that $2\pi\rho(\lambda)= (\partial_\lambda p)^\text{dr}$ and $v=(\partial_\lambda E(\lambda))^\text{dr}/(\partial_\lambda p)^\text{dr}$, we can compactly write
\be
\kappa=\frac{1}{\pi v}Z^2(\Lambda)\, .
\ee
Comparing with the Luttinger compressibility \eqref{eq_com_lutt}, we finally have
\be
K=Z^2(\Lambda)\, .
\ee
Of course, in the non interacting limit $\Delta\to 0$ case there is no dressing and $Z(\Lambda)=1$, i.e. the free fermion result as it should be.

\section{Numerical methods for the discrete inhomogeneous Luttinger liquid}
\label{app_num_discretelutt}

In this appendix we provide a numerical algorithm to compute the correlator $\langle \hat{\phi}(x)\hat{\phi}(y)\rangle$ on the ground state of the Hamiltonian \eqref{eq_dis_LL}.
Similarly to what we did in Section \ref{sub_ee_edge} in the continuum case, it is useful to define new bosonic fields
\be\label{a_def}
\hat{\varphi}_j=\frac{\hat{\phi}_j+i \hat{\Pi}_j}{\sqrt{2}}\, , \hspace{3pc} \hat{\varphi}^\dagger_j=\frac{\hat{\phi}_j-i \hat{\Pi}_j}{\sqrt{2}}\, .
\ee
Clearly $[\hat{\varphi}_j,\hat{\varphi}_{j'}]=0$ and $[\hat{\varphi}_j,\hat{\varphi}_{j'}^\dagger]=\delta_{j,j'}$. 
It is then convenient to collect the operators in an unique vector
\be\label{A_def}
\hat{\Psi}=\begin{pmatrix}\hat{\varphi}_1\\ \hat{\varphi}_2\\ ...\\ \hat{\varphi}_M\\ \hat{\varphi}_1^\dagger \\ \hat{\varphi}^\dagger_2\\...\\\hat{\varphi}_M^\dagger \end{pmatrix}\, .
\ee
Subsequently, the correlation matrix can be written as $\langle \hat{\Psi}\hat{\Psi}^\dagger\rangle$, which is a $2M\times 2M$ hermitian matrix.
Since we are ultimately interested in the correlation functions of the $\hat{\varphi}_j$ operators, it is convenient to express the  Hamiltonian in this language as well.
In particular, in this matrix-vector notation we can rephrase $\hat{H}^{\text{LL}(\ell)}$ as a bilinear form
\be
\hat{H}^{\text{LL}(\ell)}=\hat{\Psi}^\dagger \mathcal{H} \hat{\Psi}\, .
\ee
The matrix $\mathcal{H}$ has the following block form
\be
\mathcal{H}=\begin{pmatrix} \mathcal{H}^{11} && \mathcal{H}^{12} \\\mathcal{H}^{21} && \mathcal{H}^{22}\end{pmatrix}
\ee
With 
\be
\mathcal{H}^{11}_{ij}=\ell\delta_{i,j}\left(\frac{\pi K_j}{4}+\frac{1}{2\pi\ell^2(K_{j+1}+K_j)}+\frac{1}{2\pi \ell^2(K_{j-1}+K_j)}\right)-\delta_{|j-i|,1}\frac{1}{2\pi\ell(K_j+K_{i})}
\ee
\be
\mathcal{H}^{12}_{ij}=\ell\delta_{i,j}\left(-\frac{\pi K_j}{4}+\frac{1}{2\pi\ell^2(K_{j+1}+K_j)}+\frac{1}{2\pi\ell^2(K_{j-1}+K_j)}\right)-\delta_{|j-i|,1}\frac{1}{2\pi\ell(K_j+K_{i})}
\ee

Then $\mathcal{H}^{22}=\mathcal{H}^{11}$ and $\mathcal{H}^{21}=[\mathcal{H}^{12}]^\dagger$.

The Hilbert space can be regarded as a Fock space of the mode operators, which diagonalize the Hamiltonian. Let us call $\{\hat{\gamma}_j\}_{j=1,M}$ the diagonal modes and collect them in a vector $\hat{\Gamma}$ similarly to what we did with $\hat{\varphi}$ and $\hat{\Psi}$.
Then, it exists a Bogoliubov rotation implemented by a matrix $B$ connecting the two
\be
\hat{\Psi}=B\hat{\Gamma}
\ee
such that the Hamiltonian is diagonal in these modes
\be
\hat{H}^{\text{LL}(\ell)}=\hat{\Gamma}^\dagger B^\dagger \mathcal{H} B\hat{\Gamma}
\ee
where $ B^\dagger \mathcal{H} B$ is diagonal.
Crucially, the matrix $B$, in order to preserve the bosonic commutation rules, must be symplectic
\be\label{sym}
B\begin{pmatrix} \text{Id} & 0 \\ 0 &-\text{Id} \end{pmatrix} B^\dagger=\begin{pmatrix} \text{Id} & 0 \\ 0 &-\text{Id} \end{pmatrix}
\ee
Above, $\text{Id}$ is the $M\times M$ identity matrix.
Thanks to the commutation rules of the modes, we get further information on the spectrum of $\mathcal{H}$. Indeed, we must have
\be
[B^\dagger \mathcal{H} B]_{ij}=\delta_{i,j} d_j \hspace{2pc} d_j=\begin{cases} \lambda_j \hspace{1pc} & j\le n\\ \lambda_{j-n} & j>n \end{cases}
\ee
In order to have a proper well defined ground state, one must have $\lambda_j>0$. 
Assuming that $B$ has been properly chosen, the Hamiltonian in terms of the new modes looks like
\be
\hat{H}^{\text{LL}(\ell)}=\sum_j \lambda_j \hat{\gamma}^\dagger_j \hat{\gamma}_j+\lambda_j \hat{\gamma}_j \hat{\gamma}^\dagger_j\, .
\ee

In order to find the ground state, i.e. the state such that $\hat{\gamma}_j\ket{0}=0$, and its correlation functions, we rather consider a new matrix $\mathcal{G}$
\be
 \mathcal{G}=\begin{pmatrix} \text{Id} & 0 \\ 0 &-\text{Id} \end{pmatrix}B^\dagger \mathcal{H} B=B^{-1}\begin{pmatrix} \text{Id} & 0 \\ 0 &-\text{Id} \end{pmatrix} \mathcal{H} B\, ,
\ee
where in the last passage we used Eq. \eqref{sym}.
This matrix is trivially diagonal, but now notice that the $\langle \hat{\Gamma}\hat{\Gamma}^\dagger \rangle$ correlator can be expressed as a projector on the positive spectrum of $\mathcal{G}$
\be
\langle \hat{\Gamma}\hat{\Gamma}^\dagger\rangle =\sum_{\omega_j>0} \bu_{\omega_j}\bu^\dagger_{\omega_j}
\ee
where $\mathcal{G}\bu_{\omega_j}=\omega_j \bu_{\omega_j}$.
Going back to the $\langle \hat{\Psi}\hat{\Psi}^\dagger\rangle$ correlator one has
\be
\langle \hat{\Psi}\hat{\Psi}^\dagger\rangle =B\langle\hat{\Gamma} \hat{\Gamma}^\dagger \rangle B^\dagger\, .
\ee
Consider now the vectors $B\bu_{\omega_i}$. Using that the vectors $\bu_{\omega_i}$ are eigenvectors of $\mathcal{G}$ one can write
\be
\mathcal{G}\bu_{\omega_i}=\omega_i \bu_{\omega_i}\hspace{1pc}\Longrightarrow \hspace{1pc} \mathcal{H}^-\big(B\bu_{\omega_i}\big)=\lambda_i \big(B\bu_{\omega_i}\big)\, ,
\ee
where we defined $\mathcal{H}^-$ as
\be
\mathcal{H}^{-}=\begin{pmatrix} \text{Id} & 0 \\ 0 &-\text{Id} \end{pmatrix} \mathcal{H}\, .
\ee

Thus, it is convenient to introduce new vectors $\bv_{\omega_i}=B\bu_{\omega_i}$. The vectors $\bv_{\omega_i}$ are such that
\be
\mathcal{H}^-\bv_{\omega_i}=\omega_i \bv_{\omega_i}
\ee

The eigenvalue equation is not enough to unambiguously determine $\bv_{\omega_i}$. One also needs the orthogonality condition
\be\label{28}
\bv_{\omega_{j}}^\dagger\begin{pmatrix} \text{Id} & 0 \\ 0 &-\text{Id} \end{pmatrix}\bv_{\omega_i}=\bu_{\omega_{j}}^\dagger B^\dagger\begin{pmatrix} \text{Id} & 0 \\ 0 &-\text{Id} \end{pmatrix}B\bu_{\omega_i}=\bu_{\omega_{j}}^\dagger\begin{pmatrix} \text{Id} & 0 \\ 0 &-\text{Id}\end{pmatrix}\bu_{\omega_i}\, .
\ee
If $\omega_i>0$, the last term is simply $\delta_{i,j}$.
We now have all the ingredients we need to give the numerical recipe to construct the correlation matrix  $\langle \hat{\Psi} \hat{\Psi}^\dagger\rangle$, which is as it follows.
First of all, find a set of independent vectors $\mathbf{w}_i$ such that
\be
\mathcal{H}^-\mathbf{w}_i=\omega_{\mathbf{w}_i} \mathbf{w}_i\hspace{2pc} \lambda_{\mathbf{w}_i}>0\, .
\ee

The set $\{\mathbf{w}_i\}$ spans the subspace of positive eigenvalues of $\mathcal{H}^-$. No orthogonality conditions are imposed on $\mathbf{w}_i$ so far.
Subsequently, let us construct a matrix $\mathcal{O}$ whose entries are
\be
\mathcal{O}_{ij}=\mathbf{w}_i^\dagger \begin{pmatrix} \text{Id} & 0 \\ 0 &-\text{Id} \end{pmatrix}\mathbf{w}_j\, .
\ee
By definition $\mathcal{O}$ is hermitian and thus it can be diagonalized. Let $\mathcal{B}$ be the matrix of the change of basis such that $\mathcal{B}^\dagger \mathcal{O}\mathcal{B}$ is diagonal, with $\mathcal{B}$ unitary. Once in the diagonal form, we must have only positive eigenvalues according with Eq. \eqref{28}. However, we do not have the correct normalization yet.
Once we define the unnormalized vectors
\be
\tilde{\bv}_i=\sum_j \mathbf{w}_j\mathcal{B}_{ji}\, ,
\ee
we finally pose
\be
\bv_i=\left|\tilde{\bv}_i^\dagger\begin{pmatrix} \text{Id} & 0 \\ 0 &-\text{Id} \end{pmatrix}\tilde{\bv}_i\right|^{-1/2} \tilde{\bv}_i\, .
\ee
Which have now the correct normalization.
With these normalized vectors we can construct the correlation matrix as
\be
\langle \hat{\Psi} \hat{\Psi}^\dagger \rangle =\sum_i \bv_i \bv_i^\dagger\, ,
\ee
from which the discretized two point correlator of $\hat{\phi}_j$ is easily derived, according to Eq. \eqref{a_def}.

\section{The constant $C_\alpha$}
\label{app_constant_Renyi}

In this appendix we determine the constant $C_\alpha$ \eqref{eq_c_alpha}, which has been overlooked in the original reference Ref. \cite{bastianello2019renyi}.
Eq. \eqref{eq_reny_int} was computed by means of a direct computation of $\text{Tr}\hat{\rho}_I^\alpha$ using path integral methods: while doing so, constraints on the field configurations where imposed by means of Dirac's $\delta-$functions. However, notice that replacing $\delta(...)\to \text{(const.)}\times \delta(...)$ enforces the same constraint.
This ambiguity in the normalization, after taking the logarithm, results in an undetermined constant offset in the R\'enyi entropies, which can be fixed by  consistency arguments, as we are going to see. 
We start by quoting a result from Ref. \cite{bastianello2019renyi} concerning the R\'enyi entropies for open boundary conditions for a bipartition $[0,L]=I\cup \bar{I}$ where $I=[0,x_2]$ is an interval attached to the boundary. It reads
\be\label{eq_reny_int_single}
S_\alpha (0,x_2)=\frac{1}{1-\alpha}\log\left[\prod_{a=1}^{\alpha-1}\sqrt{\frac{1}{\det\left(\frac{1+\Phi^{-1}\Phi_{a/\alpha}}{2}\right)}}\, \right]+\text{const.}
\ee
Above, the definition of $\Phi_{\omega}$ is the same as in Eq. \eqref{eq_def_thetaj},the only difference being that we use the characteristic function of the interval attached to the boundary
\be
\chi_B(x)=\begin{cases}1\hspace{2pc}& x\in B=[0,x_2]\\ 0 \hspace{2pc}& \text{otherwise} \end{cases}\, .
\ee
As already mentioned, the constant offset (which does not depend on the point $x_2$, but it could depend on the integer $\alpha$) is undetermined in the method of Ref. \cite{bastianello2019renyi}, but it can be computed from the alternative technique presented in Section \ref{sub_ee_edge}, which does not suffer from any ambiguity.
A direct analytical comparison between Eq. \eqref{eq_reny_int_single} and Section \ref{sub_ee_edge} is difficult. However one can check numerically, to at least $7$ significant digits, that for arbitrary discretization step $\ell$, the constant in Eq. \eqref{eq_reny_int_single} must be zero. In the following we assume that this identity holds exactly.

We now use Eq. \eqref{eq_reny_int_single} to fix $C_\alpha$ in Eq. \eqref{eq_reny_int}. This can be done noticing that, when considering the entanglement of the bipartition $I\cup \bar{I}$ with $I=[x_1,x_2]$, if we send $x_1\to 0$ the result \eqref{eq_reny_int_single} must be recovered.
Comparing the two, one is immediately led to the conclusion
\be\label{eq:beforelimit}
C_\alpha=-\lim_{x_1\to 0}\frac{1}{1-\alpha}\log\left[\prod_{a=1}^{\alpha-1}\sqrt{\frac{1}{\pi^3\mathcal{I}_{a/\alpha}}}\sum_{\{m_j\}_{j=1}^{\alpha-1}}  \exp\Bigg\{-4\sum_{a,b=1}^{\alpha-1}\mathcal{M}_{ab}\, m_a m_b\Bigg\}\, \right]\, .
\ee
Above, $\mathcal{I}_{a/\alpha}$ and $\mathcal{M}$ are defined in Eq. \eqref{eq_M_def}. In principle, $\mathcal{I}_{a/\alpha}$ (and thus $\mathcal{M}$) still bear a non-trivial dependence on $x_2$, but the resulting expression must be $x_2-$independent.

One can also check numerically to high acccuracy that $\mathcal{I}_{a/\alpha}$ diverges in the $x_1\to 0$ limit.
The matrix elements of $\mathcal{M}$ are given by
\be
\mathcal{M}_{ab}=\sum_{c,d}^{\alpha-1}(A^\dagger)_{ac} \delta_{c,d} \alpha^{-1}\mathcal{I}^{-1}_{c/\alpha} A_{db}\, , \hspace{2pc} A_{ab}=e^{i2\pi ab/\alpha}\, .
\ee
Since the $\mathcal{I}_{a/\alpha}^{-1}$ also enters in the definition above, this means the matrix elements $\mathcal{M}_{ab}$ go to zero also. Let us now check that  (\ref{eq:beforelimit}) has a non trivial finite limit when $x_1\to 0$.  
To access it we just approximate the sum by an integral:
\be
C_\alpha=-\lim_{x_1\to 0}\frac{1}{1-\alpha}\log\left[\prod_{a=1}^{\alpha-1}\sqrt{\frac{1}{\pi^3\mathcal{I}_{a/\alpha}}}\int \dd^{\alpha-1} \tau  \exp\Bigg\{-4\sum_{a,b=1}^{\alpha-1}\mathcal{M}_{ab}\, \tau_a \tau_b\Bigg\}\, \right]\, .
\ee
The integral can be explicitly performed, leading to
\be\label{eq_D5}
\prod_{a=1}^{\alpha-1}\sqrt{\frac{1}{\pi^3\mathcal{I}_{a/\alpha}}}\int \dd^{\alpha-1} \tau  \exp\Bigg\{-4\sum_{a,b=1}^{\alpha-1}\mathcal{M}_{ab}\, \tau_a \tau_b\Bigg\}=\frac{1}{(2\pi)^{\alpha-1}}\frac{1}{\sqrt{\text{Det}\mathcal{M}}} \prod_{a=1}^{\alpha-1}\sqrt{\frac{1}{\mathcal{I}_{a/\alpha}}}\, .
\ee
The determinant in the expression above can be computed explicitly:
\be
\text{Det}\mathcal{M}=\text{Det}(A^\dagger A)\prod_{a=1}^{\alpha-1}\mathcal{I}^{-1}_{a/\alpha}\, .
\ee

The product of $\mathcal{I}_{a/\alpha}$ in the above exactly balances the product in Eq. \eqref{eq_D5}, so that the limit is, indeed, finite. The determinant $\text{Det}(A^\dagger A)$ can also be calculated:
\be
\text{Det}(A^\dagger A)=\alpha^{\alpha-2}\, .
\ee

Putting together all the terms, the value of $C_\alpha$ in Eq. \eqref{eq_c_alpha} immediately follows.

\bibliography{EE_ILL}
\end{document}